\documentclass[showpacs,twocolumn,pre,superscriptaddress]{revtex4-1}
\usepackage{mathtools}
\usepackage{epsfig}
\usepackage[utf8]{inputenc}
\usepackage{amssymb,amsthm,graphics}
\usepackage{amsmath}
\usepackage{color}
\usepackage{xspace}
\graphicspath{{Figs_SciRep/}}
\usepackage{subfig}
\usepackage[colorlinks=true,linkcolor=blue,citecolor=blue]{hyperref}
\usepackage[english]{babel}

\usepackage[T1]{fontenc} % for the polish {\k{e}}
\usepackage{multirow}
\usepackage{comment}
\usepackage{tikz}
\usetikzlibrary{decorations.pathmorphing}
\usetikzlibrary{decorations.markings}
\usetikzlibrary{arrows,shapes,snakes,automata,backgrounds,petri}
\usetikzlibrary{calc}
\usetikzlibrary{patterns}
\usetikzlibrary{decorations.text}
\tikzset{
xxtsubstrate/.style={decorate, 
line width=1pt,
draw=olive, 
decoration=snake, 
segment amplitude=0.75mm, 
line after snake=0.25mm,
line before snake=0.25mm
},
tsubstrate/.style={decorate, 
line width=1pt,
draw=olive, 
decoration=snake, 
segment amplitude=0.5mm, 
segment length=5pt,
segment amplitude=0.2mm, 
line after snake=1mm,
line before snake=1mm
},
Bsubstrate/.style={decorate, 
line width=1pt,
draw=olive, 
decoration=snake,
segment length=5pt,
segment aspect=0,
segment amplitude=0.5mm, 
line after snake=0mm,
line before snake=0mm
},
substrate/.style={decorate, 
line width=1pt,
draw=olive, 
decoration=snake, 
segment length=5pt,
segment amplitude=0.5mm, 
line after snake=0.5mm,
line before snake=0.5mm
},
activity/.style={very thick,draw=red,postaction={decorate},
decoration={markings,mark=at position .5 with
{\arrow[draw=red]{>}}}},
tactivity/.style={thick,draw=red,postaction={decorate},
decoration={markings,mark=at position .5 with
{\arrow[draw=red]{>}}}},
tEPSactivity/.style={thick,draw=red,postaction={decorate},
decoration={markings,mark=at position .55 with
{\arrow[draw=red]{>}}}},
tAactivity/.style={thick,draw=red},
Aactivity/.style={very thick,draw=red},
tSactivity/.style={thick,draw=red,postaction={decorate},
decoration={markings,mark=at position .7 with
{\arrow[draw=red]{>}}}},
Sactivity/.style={very thick,draw=red,postaction={decorate},
decoration={markings,mark=at position .7 with
{\arrow[draw=red]{>}}}}
}

\newcommand{\ai}[4]{{\left[\begin{smallmatrix}\textcolor{red}{#1}&\textcolor{red}{#2}\\\textcolor{olive}{#3}&\textcolor{olive}{#4}\end{smallmatrix}\right]}}
\newcommand{\Gammaai}[4]{\Gamma^{\ai{#1}{#2}{#3}{#4}}}
\newcommand{\Gai}[4]{G^{\ai{#1}{#2}{#3}{#4}}}

\newcommand{\Prob}{\mathcal{P}}

\newcommand{\rateextA}{e}
\newcommand{\mass}{r}
\newcommand{\rateextB}{\epsilon'}
\newcommand{\ratebranch}{s}
\newcommand{\ratehop}{H}
\newcommand{\ratedep}{\gamma}

\newcommand{\abs}[1]{\lvert #1 \rvert}

\newcommand{\dsv}{a}

\newcommand{\gcomment}[1]{}
\newcommand{\icomment}[1]{}
\newcommand{\revision}[1]{}

\newcommand{\gpvec}[1]{\mathbf{#1}}

\newcommand{\zerovec}{\gpvec{0}}
\newcommand{\nullvec}{\zerovec}

\newcommand{\dvec}{\gpvec{d}}

\newcommand{\ivec}{\gpvec{i}}
\newcommand{\jvec}{\gpvec{j}}
\newcommand{\kvec}{\gpvec{k}}

\newcommand{\xvec}{\gpvec{x}}
\newcommand{\yvec}{\gpvec{y}}

\newcommand{\plaind}{\mathrm{d}}
\newcommand{\plaine}{\mathrm{e}}
\newcommand{\plaint}{\mathrm{t}}
\newcommand{\dint}[1]{\mathchoice{\!\plaind#1\,}{\!\plaind#1\,}{\!\plaind#1\,}{\!\plaind#1\,}}
\newcommand{\ddint}[1]{\ddintx{#1}{d}}
\newcommand{\ddintx}[2]{\mathchoice{\!\plaind^{#2}#1\,}{\!\plaind^{#2}#1\,}{\!\plaind^{#2}#1\,}{\!\plaind^{#2}#1\,}}

\newcommand{\dbar}{\plaind\mkern-6mu\mathchar'26}
\newcommand{\deltabar}{\delta\mkern-8mu\mathchar'26}
\newcommand{\dintbar}[1]{\mathchoice{\!\dbar#1\,}{\!\dbar#1\,}{\!\dbar#1\,}{\!\dbar#1\,}}
\newcommand{\ddintbar}[1]{\mathchoice{\!\dbar^d#1\,}{\!\dbar^d#1\,}{\!\dbar^d#1\,}{\!\dbar^d#1\,}}

\newcommand{\Dint}[1]{\mathcal{D}#1\,}

\newcommand{\MCLineVsMathEnv}[2]{\mathchoice{#1}{#2}{#2}{#2}}

\newcommand{\ddXX}[2]{\MCLineVsMathEnv{\frac{\plaind #1}{\plaind #2}}{\plaind #1/\plaind #2}}

\newcommand{\sbraket}[1]{\lbrack #1 \rbrack}

\newcommand{\ave}[1]{\left\langle #1 \right\rangle}

\newcommand{\imag}{\mathring{\imath}}

\newcommand{\latin}[1]{{\it #1}}
\newcommand{\ie}{\latin{i.e.}\@\xspace}

\newcommand{\keyword}[1]{{\bf #1}}
\newcommand{\Adim}{\mathtt{A}}
\newcommand{\Bdim}{\mathtt{B}}
\newcommand{\Cdim}{\mathtt{C}}
\newcommand{\Ldim}{\mathtt{L}}
\newcommand{\Tdim}{\mathtt{T}}

\newcommand{\eff}{\text{\scriptsize eff}}

\begin{document}

\newcommand{\elabel}[1]{\label{eq:#1}}
\newcommand{\eref}[1]{(\ref{eq:#1})}
\newcommand{\Eref}[1]{Eq.~(\ref{eq:#1})}
\newcommand{\Erefs}[1]{Eqs.~(\ref{eq:#1})}

\newcommand{\slabel}[1]{\label{sec:#1}}
\newcommand{\srefs}[1]{Secs.~\ref{sec:#1}}
\newcommand{\sref}[1]{Sec.~\ref{sec:#1}}
\newcommand{\sre}[1]{\ref{sec:#1}}
\newcommand{\Sref}[1]{Section~\ref{sec:#1}}
\newcommand{\methodsref}[1]{methods Sec.~\ref{sec:#1}}
\newcommand{\suppsref}[1]{Sec.~\ref{sec:#1}}
\newcommand{\SuppSref}[1]{Section~\ref{sec:#1}}

\newcommand{\tlabel}[1]{\label{tab:#1}}
\newcommand{\tre}[1]{\ref{tab:#1}}
\newcommand{\tref}[1]{Tab.~\ref{tab:#1}}
\newcommand{\Tref}[1]{Table~\ref{tab:#1}}
\newcommand{\Trefs}[1]{Tabs.~\ref{tab:#1}}
\newcommand{\supptre}[1]{\ref{tab:#1}}
\newcommand{\supptref}[1]{Tab.~\ref{tab:#1}}
\newcommand{\suppTref}[1]{table~\ref{tab:#1}}
\newcommand{\supptrefs}[1]{Tabs.~\ref{tab:#1}}

\newcommand{\vlabel}[1]{\label{vid:#1}}
\newcommand{\vref}[1]{\ref{vid:#1}}
\newcommand{\Vref}[1]{Suppl. Vid.~\ref{vid:#1}}
\newcommand{\Vrefs}[1]{Suppl. Vids.~\ref{vid:#1}}

\newcommand{\flabel}[1]{\label{fig:#1}}
\newcommand{\fre}[1]{\ref{fig:#1}}
\newcommand{\fref}[1]{Fig.~\ref{fig:#1}}
\newcommand{\frefs}[1]{Figs.~\ref{fig:#1}}
\newcommand{\Fref}[1]{Figure~\ref{fig:#1}}
\newcommand{\Frefs}[1]{Figures~\ref{fig:#1}}
\newcommand{\suppfref}[1]{Fig.~\ref{fig:#1}}
\newcommand{\suppFref}[1]{Figure~\ref{fig:#1}}

\newcommand{\transpose}{\mathsf{T}}
\renewcommand{\det}[1]{\operatorname{det}\left(#1\right)}
\newcommand{\ident}{\mathbf{1}}
\newcommand{\unity}{\ident}

\newcommand{\erf}{\operatorname{erf}}
\newcommand{\erfc}{\operatorname{erfc}}

\newcommand{\AC}{\mathcal{A}}
\newcommand{\BC}{\mathcal{B}}
\newcommand{\DC}{\mathcal{D}}
\newcommand{\EC}{\mathcal{E}}
\newcommand{\FC}{\mathcal{F}}
\newcommand{\GC}{\mathcal{G}}
\newcommand{\HC}{\mathcal{H}}
\newcommand{\LC}{\mathcal{L}}
\newcommand{\NC}{\mathcal{N}}
\newcommand{\OC}{\mathcal{O}}
\newcommand{\PC}{\mathcal{P}}
\newcommand{\UC}{\mathcal{U}}
\newcommand{\VC}{\mathcal{V}}
\newcommand{\XC}{\mathcal{X}}
\newcommand{\ZC}{\mathcal{Z}}

\newcommand{\Gtilde}{\tilde{G}}
\newcommand{\GCtilde}{\tilde{\GC}}
\newcommand{\jtilde}{\tilde{j}}
\newcommand{\mtilde}{\tilde{m}}
\newcommand{\stilde}{\tilde{s}}
\newcommand{\ttilde}{\tilde{t}}
\newcommand{\utilde}{\tilde{u}}
\newcommand{\ytilde}{\tilde{y}}
\newcommand{\Gammatilde}{\tilde{\Gamma}}
\newcommand{\gammatilde}{\tilde{\gamma}}
\newcommand{\phitilde}{\tilde{\phi}}
\newcommand{\psitilde}{\tilde{\psi}}
\newcommand{\tautilde}{\tilde{\tau}}
\newcommand{\phidag}{\phi^\dag}
\newcommand{\psidag}{\psi^\dag}

\newcommand{\Ghat}{\hat{G}}
\newcommand{\What}{\hat{W}}
\newcommand{\Gammahat}{\hat{\Gamma}}
\newcommand{\phihat}{\hat{\phi}}
\newcommand{\jhat}{\hat{j}}

\newcommand{\half}{\mathchoice{\frac{1}{2}}{(1/2)}{\frac{1}{2}}{(1/2)}}
\newcommand{\threehalf}{\mathchoice{\frac{3}{2}}{(3/2)}{\frac{3}{2}}{(3/2)}}
\newcommand{\fourth}{\mathchoice{\frac{1}{4}}{(1/4)}{\frac{1}{4}}{(1/4)}}
\newcommand{\third}{\mathchoice{\frac{1}{3}}{(1/3)}{\frac{1}{3}}{(1/3)}}
\newcommand{\quarter}{\fourth}

\newcommand{\EXP}[1]{\operatorname{exp}\!\Bigg\{#1\Bigg\}}
\newcommand{\Exp}[1]{\operatorname{exp}\left(#1\right)}
\renewcommand{\exp}[1]{\mathchoice{\mathrm{e}^{#1}}{\operatorname{exp}\left(#1\right)}{\operatorname{exp}\left(#1\right)}{\operatorname{exp}\left(#1\right)}}

\title{Volume explored by a branching random walk on general graphs}
\author{Ignacio Bordeu} 
\email{ibordeu@imperial.ac.uk}
\affiliation{Department of Mathematics, Imperial College London, London SW7 2AZ, UK}
\affiliation{Centre for Complexity Science, Imperial College London, London
SW7 2AZ, UK}
\author{Saoirse Amarteifio}
\affiliation{Department of Mathematics, Imperial College London, London SW7 2AZ, UK}
\affiliation{Centre for Complexity Science, Imperial College London, London
SW7 2AZ, UK}
\author{Rosalba Garcia-Millan}
\affiliation{Department of Mathematics, Imperial College London, London SW7 2AZ, UK}
\affiliation{Centre for Complexity Science, Imperial College London, London
SW7 2AZ, UK}
\author{Benjamin Walter}
\affiliation{Department of Mathematics, Imperial College London, London SW7 2AZ, UK}
\affiliation{Centre for Complexity Science, Imperial College London, London
SW7 2AZ, UK}
\author{Nanxin Wei}
\affiliation{Department of Mathematics, Imperial College London, London SW7 2AZ, UK}
\affiliation{Centre for Complexity Science, Imperial College London, London
SW7 2AZ, UK}
\author{Gunnar Pruessner}
\email{g.pruessner@imperial.ac.uk}
\affiliation{Department of Mathematics, Imperial College London, London SW7 2AZ, UK}
\affiliation{Centre for Complexity Science, Imperial College London, London
SW7 2AZ, UK}

\begin{abstract}
Branching processes are used to model diverse social and physical scenarios, from extinction of family names to nuclear fission. However, for a better description of natural phenomena, such as viral epidemics in cellular tissues, animal populations and social networks, a spatial embedding---the branching random walk (BRW)---is required. Despite its wide range of applications, the properties of the volume explored by the BRW so far remained elusive, with exact results limited to one dimension. Here we present analytical results, supported by numerical simulations, on the scaling of the volume explored by a BRW in the critical regime, the onset of epidemics, in general environments. Our results characterise the spreading dynamics on regular lattices and general graphs, such as fractals, random trees and scale-free networks, revealing the direct relation between the graphs' dimensionality and the rate of propagation of the viral process. Furthermore, we use the BRW to determine the spectral properties of real social and metabolic networks, where we observe that a lack of information of the network structure can lead to differences in the observed behaviour of the spreading process. Our results provide observables of broad interest for the characterisation of real world lattices, tissues, and networks.
\end{abstract}

\maketitle

\noindent

Modern models of disease propagation incorporate spatial interaction
by allowing a pathogen to be passed on only to the neighbours of an infected
host \citep{EubankETAL:2004,Pastor-SatorrasETAL:2015}. A virus may multiply at a host cell and then infect any of the neighbouring ones at random \citep{Sattentau:2008}.
The total number of infected cells therefore corresponds to the number of
distinct
sites
visited by a branching random walk (BRW) \citep{DumonteilETAL:2013}, also referred to as
the Branching Wiener Sausage \citep{NekovarPruessner:2016,BerezhkovskiiMakhnovskiiSuris:1989}. In this process
active random walkers spontaneously produce descendants that carry on hopping
from site to site. At the same time, the walkers are subject to spontaneous
extinction, for example, by immune-response, healing or decay.
The average number %$\branchingRatio$ 
of descendants produced during any of these events, branching and extinction,
is known to control a transition from
a subcritical
phase,
where the disease ultimately infects only a finite number
of sites, to a supercritical phase,
where the exponential
growth of the virus eventually engulfs almost all available tissue \citep{Harris:1963}. The expected fraction
of distinct
sites visited or the size of the epidemic outbreak can be seen as the order
parameter of the process.

The characterisation of the distribution of distinct sites visited by a BRW is a long-standing problem of branching processes and random walk theory \citep{sawyer1979,DumonteilETAL:2013,RamolaMajumdarSchehr:2015}. Exact results have been obtained for one-dimensional systems~\citep{RamolaMajumdarSchehr:2015}. %Studying the volume explored by a BRW in one dimension is simplified by the property that, starting from a single infected site, every site between the leftmost and the rightmost reached sites must have been visited by an active random walker . 
However, extending such results to higher dimensional lattices and networks is met with major technical obstacles, some of which have been addressed over the past decade \citep{DumonteilETAL:2013,LeGallLin:2015,LeGallLin:2016}.

\begin{figure}[ht!]
\subfloat[\flabel{F1a_HanselnGretel}]{%
\includegraphics[width=0.47\textwidth]{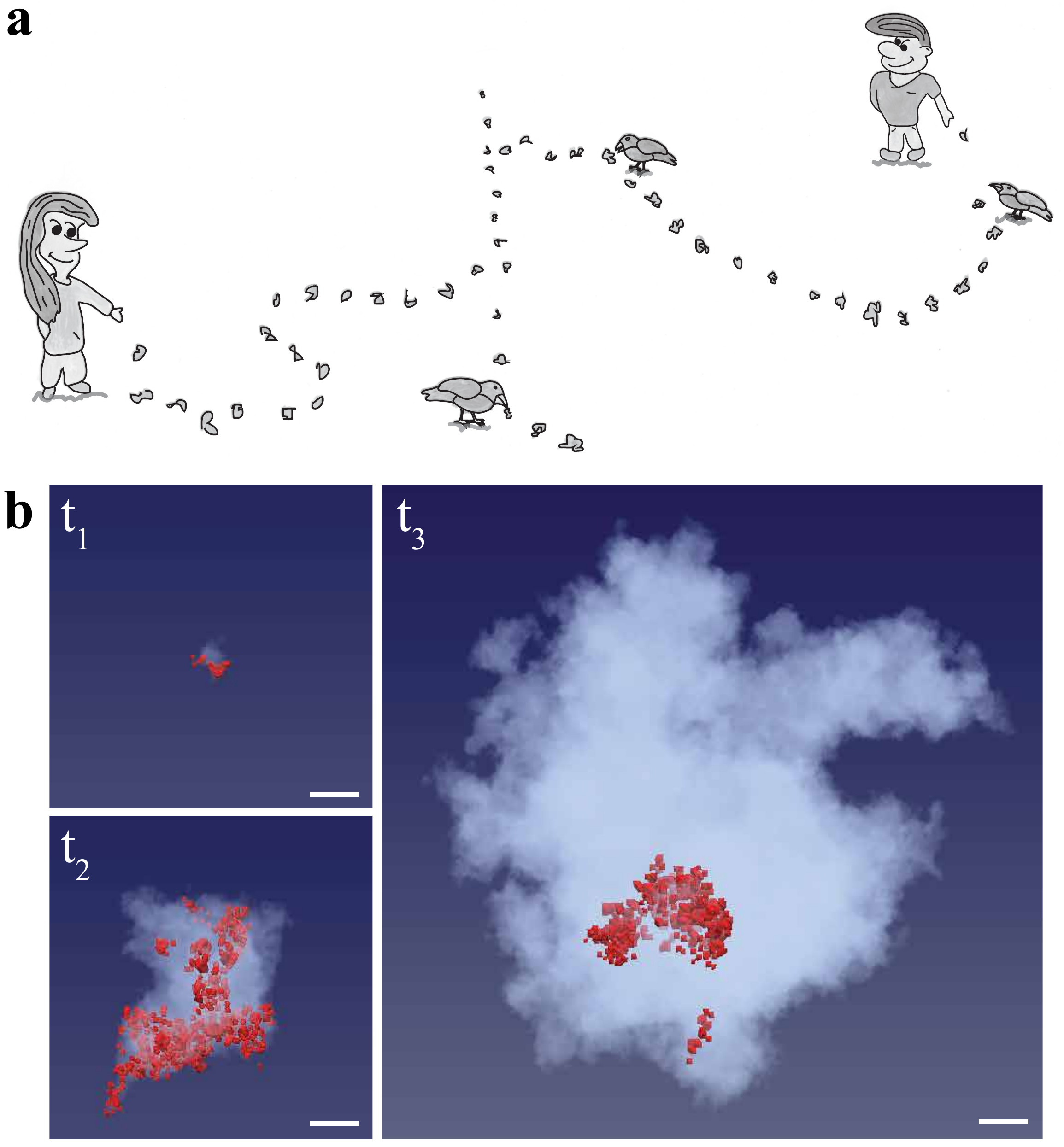}
}
\subfloat[\flabel{F1b_Cloud3d}]{%
}
\caption{{\bf Tracing the path.} {\bf a}, The active walkers, H{\"a}nsel
and Gretel, leave a trace of breadcrumbs along their way to mark the
path they have taken. Birds slowly remove the breadcrumbs, as if they were
subject to decay (regularisation, see main text). {\bf b}, Time evolution of branching random
walkers (red) and the cloud of visited sites on a 3d regular lattice at times $t_1\sim10^2$, %t=247
$t_2\sim 10^3$, %t=3932
and $t_3 \sim 10^4$. %t=62488
Scale bars are equal for all time points.}
\flabel{F01_Schematics}
\end{figure}

In the present work, we characterise analytically and, to confirm our findings, numerically the epidemic spreading in 
general graphs, including 
regular lattices, fractal, and
artificial and real complex networks, %\citep{Pastor-SatorrasETAL:2015}
at the onset of
epidemics. At this point fluctuations
are of crucial importance, dominating the dynamics.

\section{The model}
We model the epidemic as a Poisson process by considering a reaction-diffusion system of a
population of active (mobile, branching, spawning) \keyword{walkers} that hop from
their current location $\xvec$ on a graph
to any adjacent site $\yvec$ with rate $\ratehop$, and have occupation numbers $n_{\xvec}$. Walkers are further subject to
two concurrent Poisson processes, namely 
extinction with rate $\rateextA$ and
binary branching with rate $\ratebranch$, thereby producing 
\keyword{descendants}, which are indistinguishable from their ancestors.

To extract the number of distinct sites visited, we introduce an immobile
\keyword{tracer}
particle species with occupation numbers $m_{\xvec}$. They are spawned
as \keyword{offspring} by the active walkers with rate $\ratedep$ at the sites they are visiting,
thereby leaving a trail of tracers behind, similar to the breadcrumbs
left by
H{\"a}nsel and Gretel \citep{GrimmGrimm:1857}, \fref{F1a_HanselnGretel}.
We impose the constraint that
at most a single tracer can reside at any given site, which
means that the spawning of a tracer is suppressed in the presence of another tracer.
It is that suppression that generates significant complications from the point
of view of the stochastic process. 
Yet, only with this restriction in place is the number of tracers
a measure of the number of distinct sites visited by the walkers, as pictured by the {\it cloud} of visited sites in \fref{F1b_Cloud3d}.

\begin{figure}[t!]
\subfloat[\flabel{F2a_1d}]{%
\includegraphics[width=0.47\textwidth]{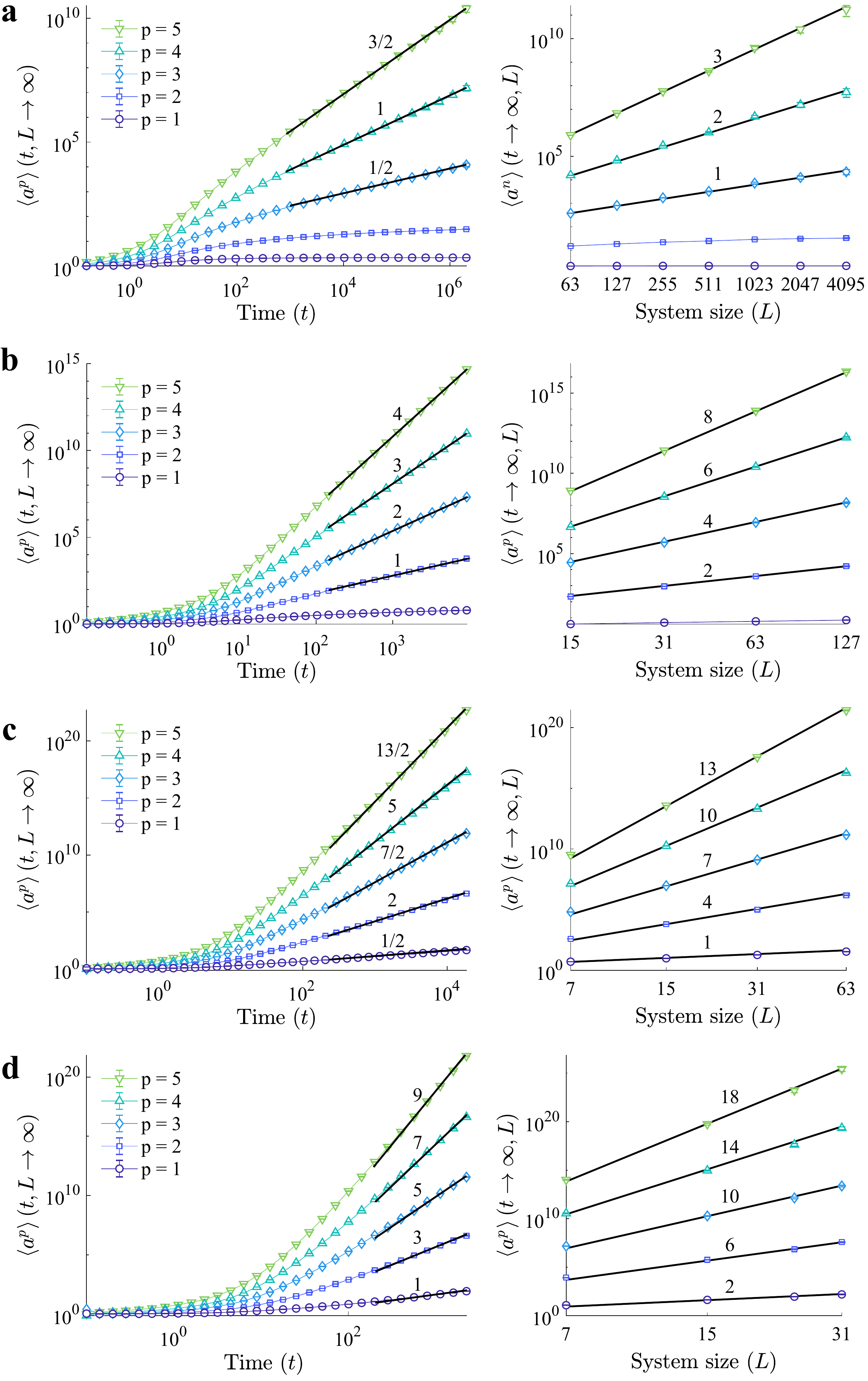}
}

\subfloat[\flabel{F2b_2d}]{%
}
\vfill
\subfloat[\flabel{F2c_3d}]{%
}
\hfill
\subfloat[\flabel{F2d_5d}]{%
}
\caption{{\bf Distinct sites visited on regular lattices.} Scaling of the moments of the numbers of distinct sites visited in time (left) and system size (right) for {\bf a}, 1d, {\bf b}, 2d, {\bf c}, 3d, and {\bf d} 5d regular lattices. Solid black lines represent the theoretical exponents given by \Eref{moment_result} for $d<4$, and \Eref{moment_result_above_dc} for $d>4$. Simulations parameters: $\ratehop = 0.1$, $\ratebranch = \rateextA = 0.45$, $\rateextB =0$, and $\ratedep \rightarrow \infty$.}
\flabel{F2_Scalings}
\end{figure}
There is no interaction between active and tracer particles, other than at the spawning of immobile tracers by active walkers.
In principle, the spawning (attempt) rate $\ratedep$ has to diverge in order
to mark every single site visited by the walkers. However, it turns out that this
limit is irrelevant as far as the asymptotic features of this process at large system sizes and long times
are concerned \citep{NekovarPruessner:2016}.

By definition,
the sets
$\{n\}$ and $\{m\}$ 
of occupation numbers $n_{\xvec}$ and $m_{\xvec}$, respectively, for each site $\xvec$ of
a given graph, are
Markovian and a master equation can be written for the joint probability 
$\Prob(\{n\},\{m\};t)$ 
to find the graph in a certain configuration of occupation numbers at
time $t$

\begin{equation}\elabel{Eq-Master_reduced}
\dot{\Prob} = \dot{\Prob}_{\ratebranch}+\dot{\Prob}_{\rateextA}+\dot{\Prob}_{\rateextB}+\dot{\Prob}_{\ratehop} +\dot{\Prob}_{\ratedep},
\end{equation}
where $\dot{\Prob}$ corresponds to the time derivative of the (joint) probability $\Prob(\{n\},\{m\};t)$, and the terms on the right-hand side, $\dot{\Prob}_\bullet=\dot{\Prob}_\bullet(\{n\},\{m\};t)$, indicate the contributions from branching $\ratebranch$, extinction of active walkers $\rateextA$ and tracer particles $\rateextB$, hopping $\ratehop$ and deposition $\ratedep$, respectively (see \suppsref{Supp_master} for details). We constructed a statistical field theory from the master equation \eref{Eq-Master_reduced} using the ladder operators introduced
by Doi \citep{Doi:1976} and Peliti \citep{Peliti:1985} (\methodsref{FT_BRW}). To regularise the propagators of the immobile particles in 
the field theory, we allow for the extinction of immobile particles with rate
$\rateextB$ in \Eref{Eq-Master_reduced}, not dissimilar
to the birds that foiled  H{\"a}nsel and Gretel's plans (\fref{F1a_HanselnGretel}). The propagators for active and tracer particles do not renormalise, and the limit $\rateextB \rightarrow 0$ is taken before any observable is evaluated. Through field-theoretic renormalisation in dimensions $d=4-\varepsilon$ we can then determine the exact scaling behaviour of the number of distinct sites visited by the walkers.

The branching process described by \Eref{Eq-Master_reduced} has three
regimes, as becomes evident in the field-theoretic formulation, where a \emph{net} extinction rate 
$\mass=\rateextA-\ratebranch$ appears. This net extinction rate is not renormalised in the field-theory and therefore no mass shift appears. The BRW is subcritical for $\mass>0$, critical for $\mass=0$ (onset of epidemics) and 
supercritical
for $\mass<0$. Hereafter, we focus on the critical case, where fluctuations dominate the dynamics, and the behaviour becomes unpredictable and highly volatile. Furthermore, for both analytical and numerical computations we consider the initial condition of a single walker at $t=0$. Extensions to different initial conditions are straight-forward.

\section{Results for regular lattices}\slabel{Results_regular}
Following the field theoretic approach (details in \srefs{FT_BRW} and \sre{FT_Renorm}) of the bulk critical behaviour in the continuum limit, where hopping is replaced by diffusion by introducing a
diffusion constant $D$, we find that in the thermodynamic limit at long times $t$, the expected number of distinct sites visited or the volume explored, $\ave{\dsv}(t,L)$, scales like $t^{(d-2)/2}$ in dimensions $d<4$.
In dimensions $d<2$ this volume remains finite 
in large $t$. 
The scaling of the $p$-th moment of the number of distinct sites visited
follows,
\begin{subequations}
\elabel{moment_result}
\begin{align}
	\ave{\dsv^p}(t,L) \propto t^{(pd-2)/2} & \quad \text{for \quad  $Dt\ll L^2$}
    \elabel{moment_result_in_t}\\
	\ave{\dsv^p}(t,L) \propto L^{(pd-2)} & \quad \text{for \quad $Dt\gg L^2$}
    \elabel{moment_result_in_L}
\end{align}
\end{subequations}
in $d<4$ provided that $pd-2>0$. 
The gap-exponent \citep{PfeutyToulouse:1977} of
$\ave{\dsv^{p+1}}/\ave{\dsv^p}$ for the scaling in $L$, which can be
thought of as the effective dimension of the cluster of visited sites,
is therefore $d$ in dimensions less than $d_c=4$.
\begin{figure}[t!]
\subfloat[\flabel{F3a_Prob_regular}]{%
\includegraphics[width=0.47\textwidth]{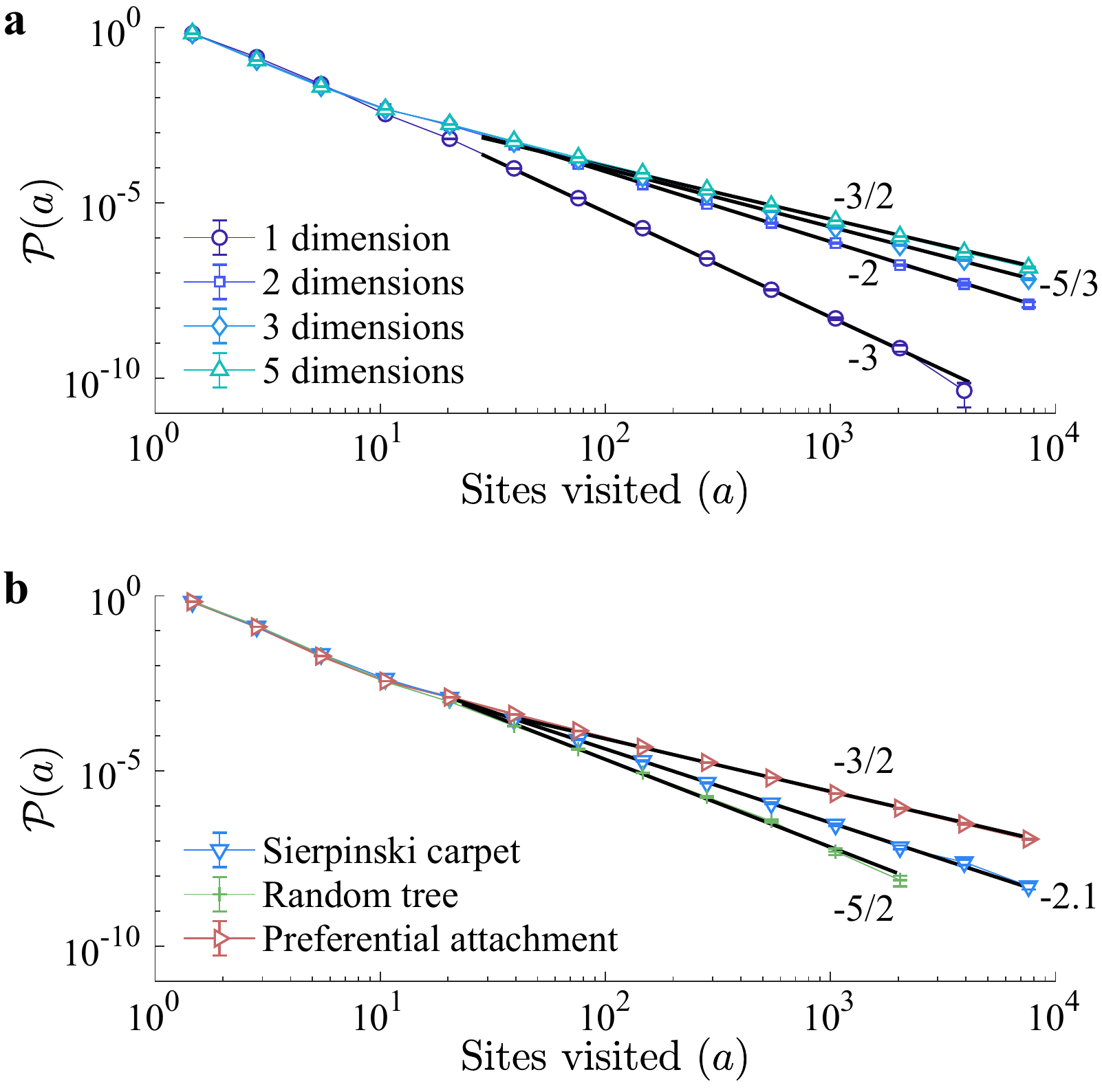}
}
\hfill
\subfloat[\flabel{F3b_Prob_graphs}]{%
}
\caption{{\bf Probability distribution of the number of visited sites}, {\bf a} for regular lattices of dimensions $d=1,2,3 \text{ and } 5$, and for {\bf b} Sierpinski carpet, random tree, and a preferential attachment (scale-free) networks. The solid black lines represent the predicted scaling given by \Eref{PDF_result}.
Simulations parameters: $\ratehop = 0.1$, $\ratebranch = \rateextA = 0.45$, $\rateextB =0$, and $\ratedep \rightarrow \infty$.}
\flabel{F3_Probs}
\end{figure}
These results describe the numerical observations  on regular lattices in
dimensions $d=1,2, \text{and } 3$ (see \sref{regular_implementation},
as shown in \frefs{F2a_1d}, \fre{F2b_2d}, and \fre{F2c_3d}, respectively, where, after an initial transient, the moments scale according to \Eref{moment_result} in time and system size (see \supptrefs{table_scaling_time} and \supptre{table_scaling_syst}).
The process is \emph{free} beyond $d_c=4$ dimensions, where
the probability 
of any walker or 
any of its ancestors or descendants ever to return to a previously visited site drops below unity, and the scaling becomes
independent of the dimension, 
\begin{subequations}
\elabel{moment_result_above_dc}
\begin{align}
	\ave{\dsv^p}(t,L) \propto t^{2p-1} &  \quad \text{for \quad $Dt\ll L^2$}
    \elabel{moment_result_above_dc_in_t}\\
	\ave{\dsv^p}(t,L) \propto L^{4p-2} & \quad \text{for \quad $Dt\gg L^2$}
    \elabel{moment_result_above_dc_in_L}
\end{align}
\end{subequations}
with logarithmic corrections in $d=d_c=4$. 
The gap-exponent in dimensions greater than $d_c=4$ is thus $4$, as confirmed by numerical observations in dimension $d=5$ (see \fref{F2d_5d} and \supptref{table_scaling_syst}).
As correlations become
irrelevant, this is usually referred to as mean-field behaviour. The set of sites visited may thus be regarded as a four-dimensional object, projected into the $d$-dimensional lattice considered.
Focusing on dimensions below $d_c=4$,
the distribution of the number
of distinct sites visited, $\dsv$,
 follows a power law, 
\begin{equation}\elabel{PDF_result}
	\PC(\dsv) = A \dsv^{-(1+2/d)} \GC(\dsv/\dsv_c)
\end{equation}
with metric factor $A$ and cutoff $\dsv_c\sim (Dt)^{d/2}$ for 
$Dt\ll L^2$ 
and 
$\dsv_c\sim L^d$ otherwise. These results show how increasing the dimensionality of the lattice promotes the appearance of larger events, evidencing the relevance of dimension on the spreading.

In dimensions $d\ge d_c=4$ the resulting scaling of the distribution is 
that of \Eref{PDF_result} at $d=4$, where the probability distribution decays like $\dsv^{-3/2}$. Numerically, we recorded, for each realisation, the total number of distinct sites visited by the process in order to construct the distribution, $\PC(\dsv)$, of sites visited. The numerical results coincide with our theoretical predictions, as shown in \fref{F3a_Prob_regular}.

\begin{figure*}[t!]
\subfloat[\flabel{F4a_SC}]{%
\includegraphics[width=0.99\textwidth]{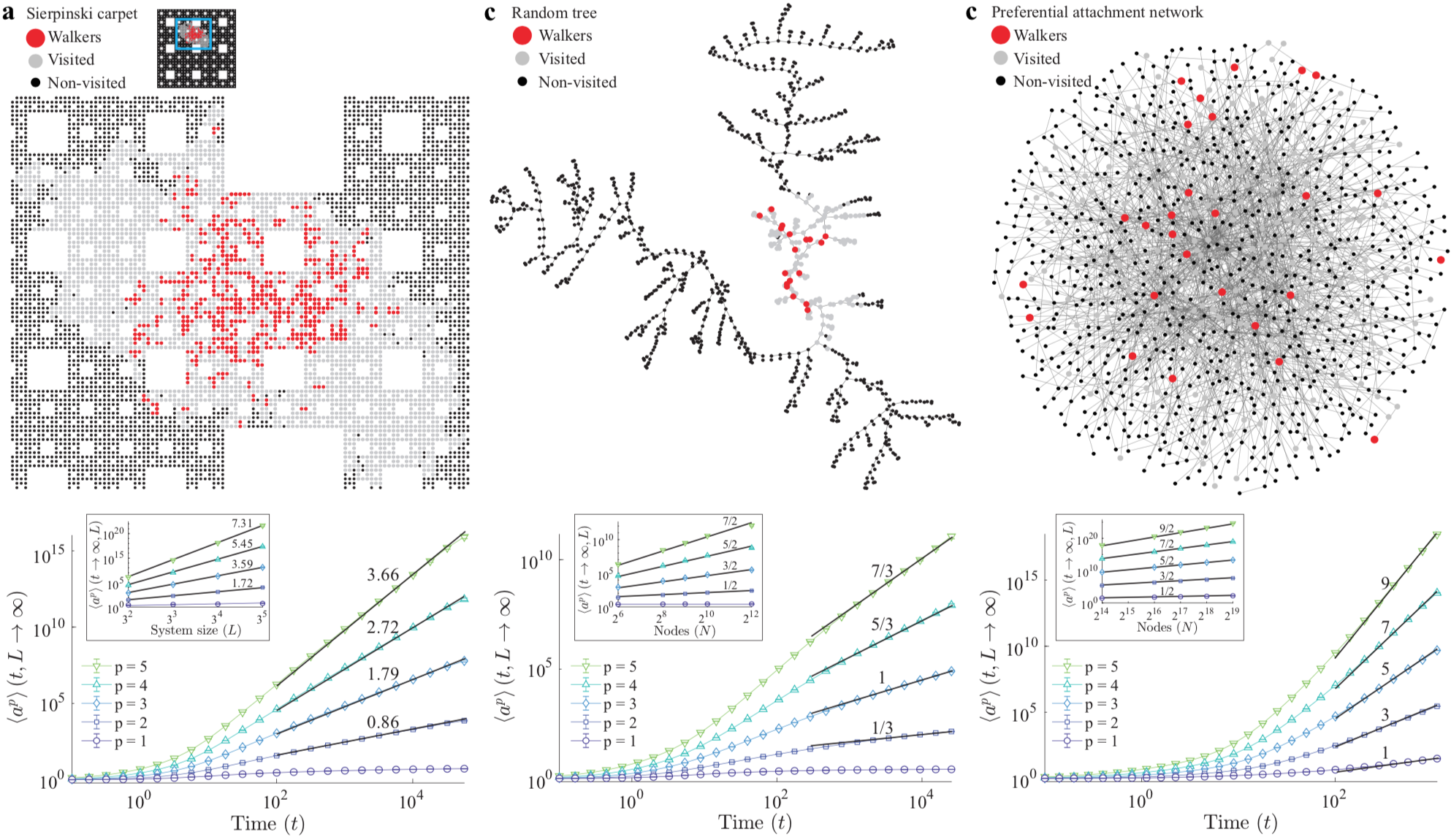}
}
\subfloat[\flabel{F4b_RT}]{%
}
\subfloat[\flabel{F4c_PA}]{%
}
\subfloat[\flabel{F4d_SC}]{%
}
\subfloat[\flabel{F4e_RT_t}]{%
}
\subfloat[\flabel{F4f_PA_t}]{%
}
\caption{{\bf Scaling on general graphs} on {\bf a}, the Sierpinski carpet, {\bf b}, random tree and, {\bf c}, preferential attachment networks. The top row shows representative states (full Sierpinski carpet shown on inset), indicating walkers (red), visited sites (grey) and non-visited sites (black). The bottom row shows the scaling of moments of the number of distinct sites visited as a function of time, and linear system size (inset), or number of nodes, in the case of networks. The solid black lines represent the predicted scaling from \Erefs{moment_result_networks}. Simulation parameters: $\ratehop = 0.1$, $\ratebranch = \rateextA = 0.45$, $\rateextB =0$, and $\ratedep \rightarrow \infty$.}
\flabel{F4_Graphs}
\end{figure*}
The exponents found above for $d=1$ are in agreement with
the exact solution by Ramola et al. \citep{RamolaMajumdarSchehr:2015}, where $\PC(\dsv)$ decays as $\dsv^{-3}$. In two dimensions the power-law tail decays as $\dsv^{-2}$, which coincides with the decay of the 2d convex hull area distribution \citep{DumonteilETAL:2013}. 

\section{Extension to general graphs}\slabel{sec_gen_graphs}
In the field theoretic approach followed to find the scaling in \sref{Results_regular} the spatial dimension of the lattice enters only in as far as its spectral dimension is concerned, which characterises the density of eigenvalues
of the Laplace operator on the graph given. 
Our results extend 
naturally to all translational invariant lattices and graphs, 
by replacing the dimension $d$ of the lattice in \Erefs{moment_result},
\eref{moment_result_above_dc} and 
\eref{PDF_result} 
by the spectral dimension $d_s$ of the graph,
as detailed in \suppsref{gen_graph}.
This holds true more generally as long as the lattice Laplacian
itself does not undergo renormalisation, \ie in the absence of an
anomalous dimension \citep{burioni2005}.
In the study of networks the number of nodes $N$, is a more natural measure of the size of the network than the linear size $L$. Using $L\sim N^{1/d_s}$ we can write the scaling of the BRW in time and number of nodes as
\begin{subequations}
\elabel{moment_result_networks}
\begin{align}
	\ave{\dsv^p}(t,N) \propto t^{(pd_s-2)/2} & \quad \text{ for \quad $Dt\ll N^{2/d_s}$}
    \elabel{moment_result_networks_in_t}\\
	\ave{\dsv^p}(t,N) \propto N^{(p-2/d_s)} & \quad \text{ for \quad $Dt\gg N^{2/d_s}$}.
    \elabel{moment_result_networks_in_L}
\end{align}
\end{subequations}
Here, the gap-exponent for the scaling in number of nodes is always unity. This extension to graphs allowed us to predict the behavior of the BRW spreading in both artificial networks relevant for social and biological sciences, and complex systems in general \citep{barabasi2000,albert2005,Pastor-SatorrasETAL:2015,wang2017}, as well as real networks. To illustrate this, we considered first the Sierpinski carpet (SCs) (\fref{F4a_SC}, \methodsref{SC_implementation}), and random trees (RTs) (\fref{F4b_RT} and \methodsref{RT_implementation}). Both of these graphs are widely applied in the context of porous media \citep{yu2001some} and percolation \citep{lyons1990random}, and have known spectral dimension: $d_s \approx 1.86$ for the SC \citep{watanabe1985spectral}, and $d_s=4/3$ for RTs \citep{destri2002}. 
Considering \Eref{moment_result} with $d=d_s$, for the SC, and \eref{moment_result_networks} for the RT we obtain accurate predictions for the spreading dynamics as confirmed by numerical simulations, \frefs{F4d_SC} and \fre{F4e_RT_t}. These theoretical predictions extend also to the distribution of visited sites (see \fref{F3b_Prob_graphs}), by setting $d=d_s$ in \eref{PDF_result}. % On RT and PA networks we observed no scaling in system size in the range $L=[2^{15},2^{20}]$ (data not shown). The networks lack of boundaries imply that the BRW can can only die off due to extinction, and does not suffer from finite-size effects.

Furthermore, we studied the BRW behaviour on a class of scale free networks \citep{barabasi1999emergence}. Since their introduction, scale free graphs have been observed to describe a plethora of natural phenomena, including the World-Wide-Web \citep{barabasi2000scale}, transportation \citep{guimera2005worldwide}, and metabolic networks \citep{jeong2000large}, to name but a few. We considered a preferential attachment scheme \citep{barabasi1999emergence} (\fref{F4c_PA}, see \methodsref{PA_implementation}), to construct networks with power-law degree distribution (\suppfref{FS2_degree_dist}).
The existence of a finite spectral gap in these networks, which indicates slow decay of return times \citep{samukhin2008laplacian,masuda2017},
suggested that the BRW process is bound to exhibit mean-field behaviour, i.e. $d_s \geq 4$. This was confirmed by numerical simulations, where the probability distribution of visited sites (\fref{F3b_Prob_graphs}) has a power-law decay with exponent $-1.52(2) \approx -3/2$, and the scaling in time and system size (\fref{F4f_PA_t} and \supptref{table_scaling_time}) follow mean-field behaviour as predicted by \eref{moment_result_networks} for $d_s=4$.

The spectral dimension gives information on the behaviour of dynamical processes on graphs. Here we use the BRW to characterise real-world networks through 
the power-law decay of the distribution of visited sites $\PC(\dsv)$, which according to \Eref{PDF_result} is $\dsv^{-(1+2/d_s)}$ provided $d_s\leq 4$. For example, the BRW exhibits near mean field-behaviour on a subset of the Facebook network, which has been characterised as scale-free \citep{viswanath-2009-activity}. Hence, we derived a large effective (spectral) dimension, $d_s=3.9(1)$, indicating a fast spreading of the viral process in this network (see \fref{prob_dist_real}).

\begin{figure}[h!]
\includegraphics[width=0.5\textwidth]{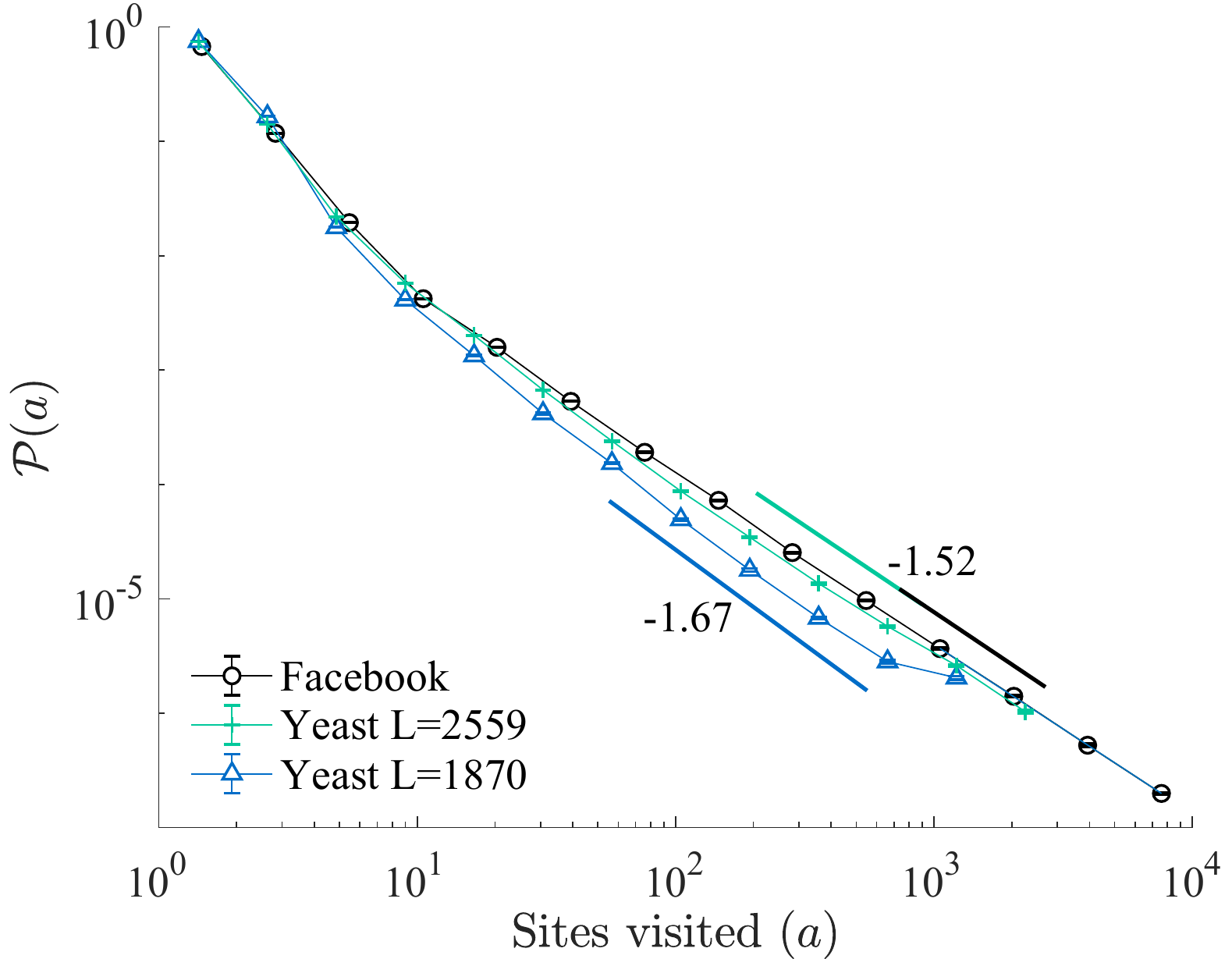}
\caption{{\bf Probability distribution of number of distinct sites visited} $\PC(\dsv)$, for the Facebook network ($L=63730$ nodes) \citep{viswanath-2009-activity}, and yeast protein interaction networks with $L=1870$ \citep{jeong2001lethality}, and $L=2559$ nodes \citep{gallos2007scaling}. %According to \Eref{PDF_result} the power-law decay of $\PC(\dsv)$ is $\dsv^{-(1+2/d_s)}$ provided $d_s\leq 4$ suggesting a spectral dimension $d_s \sim 3.83(6)$ for the large subset of the yeast protein network, $d_s = 2.97(5)$ for the small subset, and $d_s = 3.85(6) $ for the Facebook network.
The data was obtained 
from simulations of the BRW on each graph, 
with parameters $\ratehop = 0.1$, $\ratebranch = \rateextA = 0.45$, $\rateextB =0$, and $\ratedep \rightarrow \infty$.}
\flabel{prob_dist_real}
\end{figure}

We should emphasize that the spectral dimension is sensitive to changes in network topology and connectivity. To exemplify this we have considered two publicly available datasets for the yeast protein interaction network (see \fref{prob_dist_real}). We found that even though both network describe subsets of the same biochemical network, namely the complete yeast protein interactome, the spectral dimensions in both cases are significantly different, $d_s = 3.0(1)$ for the network with $N=1870$ nodes \citep{jeong2001lethality}, and $d_s = 3.8(1)$ for the larger network of $N=2559$ nodes \citep{gallos2007scaling}, leading to differences in properties of the spreading process among the two. The discrepancy points to differences in the connectivities of both networks and shows the importance of having access to the complete network in order provide a reliable analysis of their properties, which may have biological implications \citep{han2005effect,stumpf2005subnets}.

\section{Outlook}
The results presented above for the binary branching process, where walkers branch into exactly two new walkers, apply equally to more general branching processes, where the number of offspring in each birth event is given by a distribution (for details see \suppsref{k_offspring}). This can be seen, for example, in real-world scenarios where a single infected individual or device infects a whole neighbourhood around them, or in the case of signal propagation in protein networks, where the activation of one node (or chemical reaction) can activate a whole fraction of its neighboring nodes.

While the scaling behaviour
does not depend on the initial position
$\xvec_0$ of a walker, provided it is 
located in the bulk and remains there 
as the thermodynamic limit is taken,
the field theory has to be adjusted to 
account for more complicated boundary 
conditions \citep{NekovarPruessner:2016}
or the walker starting close to any such
boundary. It may also be interesting to
consider the case of initialising
each site with an independent
Poisson distributions of walkers \citep{cardy2008non}.

The approach followed in the present work provides a quantitative measure to explore and determine the
spectral dimension of artificial and real networks.
This is of particular interest when the spectral dimension is greater or equal to $2$, where the traditional approach of exploring graphs, based on simple random walks \citep{simonsen2004,masuda2017}, fails. When simulating the BRW we made the observation that 
robust scaling is more easily obtained 
on small lattices if the 
hopping rate $\ratehop$ is clearly smaller than the rates of branching $\ratebranch$ and extinction $\rateextA$. 
For large values of the hopping rate particles leave the system during the
initial transient, as seen in \fref{F2_Scalings}, thus boundary effects appear before any robust scaling can be observed. In graphs such as the PA network (\fref{F4_Graphs}), that does not have
any boundaries, these artefacts are much less pronounced.
In summary our results shed new light on the properties of spatial branching processes on general graphs, and their applicability in the study of real complex networks, and provide observables of broad interest for the characterisation of real world lattices, tissues, and networks.

\section{Methods}
\subsection{Field theory of the BRW} \slabel{FT_BRW}
In order to derive the main results for the scaling of distinct sites visited by the BRW (\sref{Results_regular}) we work along established lines \citep{TaeuberHowardVollmayr-Lee:2005}, casting the master equation in a field theory of
the annihilation fields $\phi(\xvec,t)$ and $\psi(\xvec,t)$ for the active
and the immobile
particles, respectively, and of the corresponding (Doi-shifted) creation fields 
$\phitilde(\xvec,t)$ and $\psitilde(\xvec,t)$. The governing 
Liouvillian $\LC=\LC_0+\LC_1$ consists of a harmonic part,
\begin{multline}\elabel{Eq_Harmonic}
\LC_0(\phi,\psi,\phitilde,\psitilde)
= -\phitilde\partial_t\phi
+ D\phitilde\nabla^2\phi 
- \mass \phitilde\phi\\
- \psitilde\partial_t\psi 
- \rateextB \psitilde\psi
+ \tau\psitilde\phi,
\end{multline}
and a non-linear part,     
\begin{multline}\elabel{Eq_Nonlinear}
\LC_1(\phi,\psi,\phitilde,\psitilde)
= s\phitilde^2\phi 
+ \sigma\psitilde\phitilde\phi
- \lambda\psitilde\psi\phi\\
- \xi\psitilde^2\psi\phitilde\phi
- \kappa\psitilde\psi\phitilde\phi
- \chi\psitilde^2\psi\phi,
\end{multline}
where we have taken the continuum limit. The space and time integrated Liouvillian produces the field-theoretic action $\mathcal{A}=\int\ddint{x}\dint{t} \LC$, whose exponential $\plaine^\mathcal{A}$ enters into the path integral formulation.
The couplings in the Liouvillian are related to the rates in the master equation as follows:
$D$ is a diffusion constant $D=\ratehop\Delta x^2$, where $\Delta x$ is the lattice spacing, and $\ratehop\propto\Delta x^{-2}$ when
the limit $\Delta x\to0$ is taken, in order to maintain finite diffusivity. At bare level the non-linear couplings, with the exception of the branching rate $s$, are equal to spawning rate $\gamma$, i.e. $\tau=\sigma=\lambda=\xi=\kappa=\chi=\gamma$. This follows from translating the master \Eref{Eq-Master_reduced} into field-theoretic language (see \sref{Supp_master} for details).

At
the same time the \emph{net} extinction rate 
$\mass=\rateextA-\ratebranch$, the field-theoretic mass of the walkers,
has to be kept finite. In this parameterisation,
there are three regimes, as described in the main text: a
subcritical one for $\mass>0$, a critical for $\mass=0$ and a 
supercritical
for $\mass<0$.
In the field theory, all large scale (infrared) phenomena will be controlled
by $\mass\to0^+$, which corresponds to the onset of epidemics, the limit
studied in this work.
The mass of the tracers, $\rateextB$, serves merely
as a regularisation, and is removed by taking the limit $\rateextB\to0$. The bare transmutation rate $\tau$, corresponding to $\ratedep$
on the lattice, and the bare branching rate $s$ 
of the active particles ($\ratebranch$ on the lattice) are the two processes
that we expect will govern
all infrared behaviour in all dimensions and are therefore assumed to be dimensionless.
These two choices determine the engineering dimension %\citep{Amit_or_LeBellac_or_Taeuber}
\citep{Taeuber:2014} of all other bare couplings,
resulting in $\xi$, $\kappa$ and $\chi$ being infrared irrelevant.
Together with $\lambda$, these four couplings are due to the suppression of the
spawning of tracers when a site is occupied already. At the upper critical
dimension, $d_c=4$, the coupling $\lambda$ 
is marginally relevant, being infrared irrelevant above and relevant below. The 
minimal subtraction scheme %\cite{Amit_or_LeBellac_or_Taeuber}
 \citep{Taeuber:2014} we have used will produce results in terms
of $\varepsilon=4-d$.

The Liouvillian constructed above is the object that allows the exact calculation of the scaling exponents of the $p$-th moment of the volume explored by a branching random walk $\ave{\dsv^p}(t,L)$, in time $t$, and linear system size $L$. Initialising the system at time $t_0=0$ with a single active walker at position $\xvec_0$, field-theoretically implemented by the creation field $\phitilde(\xvec_0,0)$, the ensemble average $\ave{\dsv}(t,L)$ of the volume explored by the BRW is determined by
\begin{equation}\elabel{first_moments}
	\ave{\dsv}(t,L) =
	\int \ddint{x}
	\ave{\psi(\xvec,t)\phitilde(\xvec_0,0)},
\end{equation}
where the density of tracers particles at position $\xvec$ and time $t>0$ is measured by $\psi(\xvec,t)$ and integrated over all space. Similarly \citep{NekovarPruessner:2016}, higher moments are determined
by integrals of the form
\begin{multline}\elabel{dsv_moments}
	\ave{\dsv^p}(t,L) 
    \\ =
	\int \ddint{x_p} \ldots \ddint{x_1}
	\ave{\psi(\xvec_p,t)\ldots\psi(\xvec_1,t)\phitilde(\xvec_0,0)},
\end{multline}
or equivalently, by evaluating the Fourier transform at spatial momentum $\kvec = 0$. These are functions of the couplings introduced above, but to leading order not of the walker's initial position $\xvec_0$, provided it is located in the bulk. We implement this numerically by always placing the walker initially at the centre site of odd-sized regular lattices, see \sref{regular_implementation}. The average $\ave{\bullet}$ introduced on the right hand side of \Eref{first_moments} correspond to the path integral
\begin{multline}\elabel{observable_example}
	\ave{\psi(\xvec_p,t)\ldots\psi(\xvec_1,t)\phitilde(\xvec_0,0)}
	\\ =
	\int \Dint\Pi
	\big( \psi(\xvec_p,t)\ldots\psi(\xvec_1,t)\phitilde(\xvec_0,0) \big)
	\exp{\int\ddint{x}\dint{t} \LC}\ ,
\end{multline}
which measures the $p$-point correlation function of tracers at
$(\xvec_i,t)$, $i=1,2,\ldots, p$ in
response to the 
creation of a walker at $(\xvec_0,t=0)$. Here, the integration measure is $\Dint\Pi = \Dint\phi\Dint\phitilde\Dint\psi\Dint\psitilde$. Field theoretic renormalisation in dimensions $d=4-\varepsilon$ then allows us to derive the scaling of the number of distinct sites visited (see \suppsref{FT_Renorm} for more details). 

\subsection{Numerical implementation}\slabel{implementation}
In the numerical implementation,
an active particle is allowed to diffuse by hopping from the site it resides on 
to a nearest neighbouring site with rate $\ratehop$, 
 branch with rate $\ratebranch$ by placing an identical offspring at the present site 
or become extinct with rate $\rateextA$. Each distinct site visited is recorded, equivalent to taking the limit $\ratedep\rightarrow\infty$ in the theory. The instantaneous number $\dsv(t,L)$ 
of distinct sites visited up to time $t$ is therefore the number of sites recorded.
Parameters were chosen such that $\ratehop+\ratebranch+\rateextA=1$, $\ratehop$ was set to $0.1$, and $\rateextA=\ratebranch = 0.45$.
If $M$ walkers are present in the system at a given time the waiting time for the next event (hopping, branching or extinction) is determined by 
$-\ln(1-u)/M$ where $[0,1)\ni u\sim U(0,1)$ is a uniformly distributed random variable. 
For every lattice size we performed $10^6-10^9$ realisations of the process. 

\subsubsection{Regular, integer-dimensional lattices}\slabel{regular_implementation}
The regular lattices studied here are hypercubic $d$-dimensional lattices, characterised by their linear size $L=2^m-1$, $m \geq 4$, which is chosen to be odd so that 
it contains a well-defined centre site, on which the single active walker is initially placed. %\revision{A different choice for the position of the initial walker does not affect the observed scalings}.
To study finite-size scaling, absorbing boundary conditions were applied. However, we observed that the boundary conditions have no effect on the scaling (data not shown). 
The numerical results were fitted to a power-law as described in \sref{fitting}, to obtain the values in \supptrefs{table_scaling_time} and \supptre{table_scaling_syst}.

\subsubsection{Sierpinski carpet} \slabel{SC_implementation}
The Sierpinski carpets were constructed from two dimensional lattices of linear dimension $3^m$, $m\geq 2$. The lattice was divided into $3^2$ equal sub-squares each of size $3^{m-1}$, the central square was removed, leaving $3^2-1$ sub-squares. The procedure is iterated over the remaining sub-squares. The
spectral dimension of the Sierpinski carpet has been estimated to be $d_s= 1.86$ \citep{watanabe1985spectral,dasgupta1999scaling}.
A random point around the central hole of the fractal was used as the initial location of the walker in every realisation. 

\subsubsection{Random trees} \slabel{RT_implementation}
 The critical random tree networks \citep{destri2002growth} were constructed as a critical Galton-Watson process, where every node has either 0, 1, or 2 descendants, such that the mean degree of the network is 2. We generated networks with $2^{6}-2^{12}$ nodes. These graphs have no closed loops. 
The spectral dimension of the random tree ensemble is $d_s=4/3$ 
\citep{destri2002}.
For every realisation of the process, a new random tree was generated, and a node was selected at random as the starting location of the initial walker.
\subsubsection{Preferential-attachment network} \slabel{PA_implementation}
A preferential attachment (PA) network is a class of scale-free networks, characterised by a power-law degree distribution.
We followed the Barabási–Albert model of preferential attachment \cite{barabasi1999emergence} initialised with a single node to generate networks with $2^{12}-2^{19}$ nodes. The networks have power-law degree distribution with exponent $-2.9$ and mean degree $\ave{k_i}=6.3$ (see \suppfref{FS2_degree_dist}). For every realisation of the process, a new network was constructed, and a node was selected at random as the starting location of the initial walker.

\acknowledgments
We thank Aman Pujara, Brandon Annesi, 
Kin Tat Yiu,
Henry Grieve
and Henry Wilkes
for their contributions at an earlier stage of this project. 
We thank Renaud Lambiotte and Erwin Frey for their useful comments. 
We would also like
to thank Andy Thomas and Nemar Porats for their tireless computing support.
I.B. acknowledges the support of CONICYT PhD scholarship (Chile), Beca de Doctorado en el Extranjero No. 72160465, and the Centre for Doctoral Training on Theory and Simulation of Materials at Imperial College London, EPSRC (EP/L015579/1).

\section*{Author contributions}
I.B. and G.P. wrote the paper. 
S.A. wrote the algorithms for lattices of dimensions 2, 3, and 5, and networks. R.G-M. wrote the algorithm for the 1 dimensional lattice. R.G-M. and B.W. wrote the fitting routines. S.A., R.G-M., B.W. and I.B. analysed the numerical results with support of G.P. N.W. contributed to the extension of the theory to general graphs. All authors worked on the theory, read and approved the manuscript.

\section*{Competing interests}
The authors declare no competing interests.

\section*{Data Availability}
The numerical data is available upon request, by contacting I.B., ibordeu@imperial.ac.uk.

\section*{Additional information}
Supplementary information can be found online. Correspondence should be addressed to I.B. or G.P.

\onecolumngrid
\pagebreak

\setcounter{section}{0}
\setcounter{equation}{0}
\setcounter{figure}{0}
\setcounter{table}{0}
\setcounter{page}{1}
\makeatletter
\renewcommand{\thesection}{S\arabic{section}}
\renewcommand{\theequation}{S\arabic{equation}}
\renewcommand{\thetable}{S\arabic{table}}
\renewcommand{\thefigure}{S\arabic{figure}}
\renewcommand{\bibnumfmt}[1]{[S#1]}
\renewcommand{\citenumfont}[1]{S#1}

\vspace*{10mm}
\noindent
\textbf{\Large{Supplementary material}} \\
\vspace*{3mm}

\noindent
\textbf{\huge{Volume explored by a branching random walk on general graphs}}\\
\vspace*{3mm}

\noindent
\textbf{\large{Ignacio Bordeu}}$^{1,2,\text{*}}$\textbf{\large{, Saoirse Amarteifio}}$^{1, 2}$\textbf{\large{, Rosalba Garcia-Millan}}$^{1,2}$\textbf{\large{, Benjamin Walter}}$^{1,2}$\textbf{\large{, Nanxin Wei}}$^{1,2}$\textbf{\large{, and Gunnar Pruessner}}$^{1,2,\dag}$\\
\vspace*{3mm}

\noindent
$^1$Department of Mathematics, Imperial College London, London SW7 2AZ, UK\\
\noindent
$^2$Centre for Complexity Science, Imperial College London, London SW7 2AZ, UK\\
\noindent
$^*$corresponding author: ibordeu@imperial.ac.uk\\
$^\dag$g.pruessner@imperial.ac.uk
\vfill

\pagebreak

\section{Master equation for the branching random walk} \slabel{Supp_master}
In the following we describe the contributions to the master equation \eref{Eq-Master_reduced} from each of the processes the branching random walk comprises. The contribution to the master equation for the joint probability 
$\Prob(\{n\},\{m\};t)$ from
the spawning of immobile tracer particles by active walkers must take into account the finite carrying capacity $\bar{m}_0$ of each lattice site. To account for a finite carrying capacity an effective deposition rate is introduced that decays linearly with the number of tracer particles already present at the site of interest \citep{SNekovarPruessner:2016},
\begin{equation}
    \ratedep_{\text{eff}} = \ratedep \frac{\bar{m}_0-m_\xvec}{\bar{m}_0}.\nonumber
\end{equation}
To study the number of distinct sites visited $\bar{m}_0$ is set to 1. With this constraint in place, each site visited is {\it marked} with a tracer particle at most once, so that their total number 
is that of \emph{distinct} sites visited by the BRW. %The effect of considering different carrying capacities is discussed later. 
With these considerations, the contributions to the master equation from deposition of tracer particles read

\begin{equation}\elabel{Eq-MasterB}
\dot{\Prob}_{\ratedep}(\{n\},\{m\};t) = 
\ratedep
\sum_{\xvec}
\Big((1-(m_{\xvec}-1)) n_{\xvec} \Prob(\{n\},\{\ldots,m_{\xvec}-1,\ldots\};t)
-
(1 - m_{\xvec}) n_{\xvec} \Prob(\{n\},\{m\};t)\Big),
\end{equation}
where $n_{\xvec}$ and $m_{\xvec}$ correspond to the number of active and immobile particles at site ${\xvec}$. The sum $\sum_{\xvec}$ runs over all lattice sites. The contribution from branching of active walkers reads
\begin{equation}\elabel{ME_branching}
\dot{\Prob}_{\ratebranch}(\{n\},\{m\};t) = 
\ratebranch
\sum_{\xvec}
 \Big(  ((n_{\xvec}-1) \Prob(\{\ldots,n_{\xvec}-1,\ldots\},\{m\};t)
                - n_{\xvec} \Prob(\{n\},\{m\};t) \Big).
\end{equation}
The contributions from extinction are
\begin{equation}\elabel{ME_extA}
\dot{\Prob}_{\rateextA}(\{n\},\{m\};t) = 
\rateextA
\sum_{\xvec}
\Big(   (n_{\xvec}+1) \Prob(\{\ldots,n_{\xvec}+1,\ldots\},\{m\};t)
                - n_{\xvec} \Prob(\{n\},\{m\};t) \Big)
\end{equation}
for active particles, and
\begin{equation}
\dot{\Prob}_{\rateextB}(\{n\},\{m\};t) = 
\rateextB
\sum_{\xvec}
 \Big(   (m_{\xvec}+1) \Prob(\{n\},\{\ldots,m_{\xvec}+1,\ldots\};t)
                - m_{\xvec} \Prob(\{n\},\{m\};t) \Big)
\end{equation}
for immobile particles. Finally, the contribution to the joint probability from the hopping of active walkers reads 
\begin{equation}
\dot{\Prob}_{\ratehop}(\{n\},\{m\};t) = 
\frac{\ratehop}{q}
\sum_{\xvec}
 \sum_{\yvec.nn.\xvec}
          \Big(   (n_{\yvec}+1) \Prob(\{\ldots,n_{\xvec}-1,\ldots,n_{\yvec}+1,\ldots\},\{m\};t)
                - n_{\xvec} \Prob(\{n\},\{m\};t) \Big),
\end{equation}
where the sum $\sum_{\yvec.nn.\xvec}$ runs over all $q$ nearest neighbouring ({\it nn}) sites
$\yvec$ of $\xvec$. 

Combining the contributions from all the subprocesses, the master equation for the joint probability $\Prob=\Prob(\{n\},\{m\};t)$ reads
\begin{equation}
\dot{\Prob} = \dot{\Prob}_{\ratebranch}+\dot{\Prob}_{\rateextA}+\dot{\Prob}_{\rateextB}+\dot{\Prob}_{\ratehop} +\dot{\Prob}_{\ratedep}.
\end{equation}
as shown in \Eref{Eq-Master_reduced}.

\section{Field-theory of the BRW} \slabel{FT_Renorm}
In the following, we show the details of the field-theoretical calculations performed to obtain the main results of the article, \Erefs{moment_result}, \eref{PDF_result}, and \eref{moment_result_above_dc}. In
\suppsref{Sec_Dim_An_Couplings} we describe the dimensional analysis of the bare couplings. In \suppsref{propagarots}, we introduce a diagrammatic representation of the propagators and couplings, and in \suppsref{Sec-RelevantInter} we determine the relevant interactions. In \suppsref{renorm}, we perform the renormalisation of the couplings, and finally calculate the higher order correlations that give rise to the scaling of the moments of the number of distinct sites visited in \suppsref{Callan_Sym}.

\subsection{Dimensional analysis of the bare couplings}
\slabel{Sec_Dim_An_Couplings}
To compute the critical dimension of the process described by the Liouvillian $\LC=\LC_0+\LC_1$, \Erefs{Eq_Harmonic} and \eref{Eq_Nonlinear}, and to extract the relevant interactions, i.e the couplings that remain relevant in every spatial dimension, we study the engineering dimensions (here, represented by $\sbraket{\cdot}$) of every coupling in the action. We expect that the long range physics in time and space is governed by three processes: diffusion with constant $D$, branching with rate $\ratebranch$, and transmutation with rate $\tau$. Introducing three independent dimensions, namely $\Adim$, $\Bdim$ and $\Cdim$, we impose 
\begin{equation}\elabel{dimana_basics}
\sbraket{\tau} = \Adim, \quad 
\sbraket{s} = \Bdim, \quad \text{and} \quad
\sbraket{D} = \Cdim.
\end{equation}
With $\sbraket{\xvec}=\Ldim$, $\sbraket{t} = \Tdim$, and $\sbraket{\partial_t}=\sbraket{D\nabla^2}$ it follows that $\Tdim = \Cdim\Ldim^2$ is not an independent dimension. As the action, $\mathcal{A}=\int\ddint{x}\dint{t} \LC$, itself must be dimensionless, i.e. $\sbraket{\mathcal{A}} = 1$, we obtain $\sbraket{\mass} = \Tdim^{-1}=\Cdim\Ldim^{-2}$ and

\begin{equation}
\left[ \phitilde \right] = \Bdim^{-1}\Cdim\Ldim^{-2}, \qquad
\left[ \phi \right] = \Bdim \Cdim^{-1} \Ldim^{2-d}, \qquad
\left[ \psitilde \right] = \Adim^{-1}\Bdim^{-1}\Cdim^2\Tdim^{-2}, \qquad
\left[ \psi \right] = \Adim\Bdim\Cdim^{-2}\Ldim^{4-d}
\end{equation}
for the fields in real time and space, such that $[\phitilde\phi] = [\psitilde\psi] = \Ldim^{-d}$. The engineering dimensions of the couplings follow: 

\begin{subequations}\elabel{dimana_inter}
\begin{equation}
\sbraket{\lambda} = \Bdim^{-1}\Cdim^2\Ldim^{d-4} \qquad
\sbraket{\sigma} = \Adim\Bdim\Cdim^{-1}\Ldim^2 \qquad
\sbraket{\chi} = \Adim\Ldim^{d} 
\end{equation}

\begin{equation}
\sbraket{\kappa} = \Cdim\Ldim^{d-2} \qquad
\sbraket{\xi} = \Adim\Bdim\Cdim^{-1}\Ldim^{d+2}.
\end{equation}
\end{subequations}

Setting $\Adim = \Bdim = \Cdim = 1$, we find a critical dimension $d_c = 4$, above which all of these interactions become irrelevant. At the critical dimension $d=d_c=4$ the couplings $\sigma$, $\chi$, $\kappa$, and $\xi$ remain irrelevant, while $\lambda$ becomes marginal. To regularise the ultraviolet we work in dimensions $d=4-\epsilon<4$.

As a point of discussion, we note that
other choices of independent dimensions are
possible, limited only by the symmetries to be preserved. 
Initially we considered $\sigma$, rather than $\tau$ to have an independent dimension. The resulting (very messy) field theory depends on the non-universal, bare value of $s$ and 
produces 
no renormalisation of $\tau$, which, however, \emph{must} renormalise as 
$\ave{\dsv}(t,L)\sim \tau_{\eff} L^2$
(see \Sref{renorm}) and cannot scale faster than
the volume of the system, $L^d$.

A coupling with independent dimension is saved from changing
relevance and thus from 
possible irrelevance
in the infrared limit of large space and long time. 
The choice of dimensions is therefore a
choice of interactions that
ultimately govern the infrared. 
If the stochastic process under consideration takes
place on the lattice,
this may be determined by taking
the continuum limit, provided the process 
does not possess any competing
scales, in which case the continuum 
limit coincides with the 
thermodynamic limit of infinite
system size. However, as soon as different
processes and scales compete, 
such as hopping, branching, spawning and
extinction rates in the present case,
the continuum
limit is a mere approximation of the original
process on the lattice and the choice
of (independent) dimensions becomes a
claim about which interactions govern the 
infrared. For example, 
considering a biased random walk,
letting space scale
linear in time preserves a drift but
removes diffusion,
while letting space scale quadratically
in time preserves the latter, while the 
drift velocity diverges.

\subsection{Fourier transform} \slabel{Fourier}
Throughout the manuscript, we denote the Fourier transform $\mathcal{F}[f(\xvec,t)]$ of a function $f(\xvec,t)$ in space $\xvec$ and time $t$ simply as $f(\kvec,\omega)$, where the spatial momentum $\kvec$ is the conjugate of the position $\xvec$, and the frequency $\omega$ is the conjugate of time $t$. The direct Fourier transform is defined as

\begin{equation}
f(\kvec,\omega) = \int \exp{\imag\omega t-\imag\kvec \cdot\xvec}f(\xvec,t) \plaind^d\xvec \plaind\plaint ,
\end{equation}
so that the inverse Fourier transform is
\begin{equation}
f(\xvec,t) = \int \exp{-\imag\omega t+\imag\kvec\cdot\xvec}f(\kvec,\omega) \dbar^d\kvec\dbar\omega ,
\end{equation}
where $\dbar^d\kvec = (1/2\pi)^d \plaind\kvec$, $\dbar\omega = (1/2\pi)\plaind\omega$, and $d$ is the spatial dimension.

\subsection{Propagators and couplings}\slabel{propagarots}
We begin by considering the field-theoretic action $\mathcal{A}=-\int\ddint{x}\dint{t} \LC$, where the terms in the
Liouvillian $\LC=\LC_0+\LC_1$ are given by \Erefs{Eq_Harmonic} and \eref{Eq_Nonlinear}, respectively. 
In order to render the Laplacian term local
the action is rewritten in Fourier space,
where the momentum $\kvec$ is the conjugate of position $\xvec$ and 
the frequency $\omega$ is the conjugate of time $t$ (as defined in \suppsref{Fourier}).
The perturbative renormalisation scheme starts by reading off the propagators
from the bilinear part, introducing a diagrammatic language as we proceed.
For the walkers the bare propagator reads
\begin{equation}\elabel{PropA}
\ave{\phi(\kvec,\omega)\phitilde(\kvec',\omega')}_0 = 
\frac{\deltabar(\kvec+\kvec')\deltabar(\omega+\omega')}{-\imag\omega+Dk^2+\mass}\corresponds \tikz
[baseline=-2.5pt]{
\draw[Aactivity] (0.5,0) -- (-0.5,0) node[at end,above] {};
},
\end{equation}
where $\deltabar(\kvec+\kvec')=(2\pi)^d\delta(\kvec+\kvec')$ denotes a scaled
$d$-dimensional Dirac-$\delta$ function, and correspondingly for 
$\deltabar(\omega+\omega')$. Diagrammatically, the bare propagator is shown as a straight line.
For the tracers the bare propagator becomes
\begin{equation}\elabel{PropB}
\ave{\psi(\kvec,\omega)\psitilde(\kvec',\omega')}_0 = 
\frac{\deltabar(\kvec+\kvec')\deltabar(\omega+\omega')}{-\imag\omega+\rateextB}\corresponds
\tikz[baseline=-2.5pt]{
\draw[substrate] (0.5,0) -- (-0.5,0) node[at end,above] {};
}\ ,
\end{equation}
diagrammatically shown as a wavy line.
Both bare propagators carry a positive mass, $\mass=\rateextA-\ratebranch$ in \Eref{PropA} and $\rateextB$ in \Eref{PropB}
which guarantees causality as the inverse Fourier transform will generate a Heaviside-$\theta$ function 
in time. Both propagators \Erefs{PropA} and \eref{PropB} do not undergo renormalisation.
Finally, the transmutation vertex features in
\begin{equation}\elabel{tau_vertex}
\ave{\psi(\kvec,\omega)\phitilde(\kvec',\omega')}_0 = 
\tau \frac{\deltabar(\kvec+\kvec')\deltabar(\omega+\omega')}{(-\imag\omega+\rateextB)(-\imag\omega +Dk^2+\mass)}
\corresponds
\tikz[baseline=-2.5pt]{
\draw[Aactivity] (0.5,0) -- (0,0) node[at end,above] {$\tau$};
\draw[substrate] (0,0) -- (-0.5,0) node[midway,above] {};
}
\end{equation}
and signals the appearance of a tracer in response to the presence of a walker, as time is to be read from right to left.
The non-linear part of the Liouvillian, $\LC_1$, contributes with six interaction vertices, which diagrammatically read 

\begin{eqnarray}\label{mu_kappa}
\tikz[baseline=-2.5pt]{
\draw[Aactivity] (0.5,0) -- (0,0) node[at end,above] {$s$};
\draw[Aactivity] (130:0.5) -- (0,0);
\draw[Aactivity] (-130:0.5) -- (0,0);
}
&\qquad&
\tikz[baseline=-2.5pt]{
\draw[Aactivity] (0.5,0) -- (0,0) node[at end,above] {$\sigma$};
\draw[Aactivity] (0,0) -- (-0.5,0) node[at end,above] {};
\draw[substrate] (-130:0.5) -- (0,0);
}\\
\tikz[baseline=-2.5pt]{
\draw[Aactivity] (0.5,0) -- (0,0) node[at end,above] {$\xi$};
\draw[Aactivity] (0,0) -- (-0.5,0) node[at end,above] {};
\draw[substrate] (130:0.5) -- (0,0);
\draw[substrate] (-130:0.5) -- (0,0);
\draw[substrate] (-50:0.5) -- (0,0);
}
&\qquad&
\tikz[baseline=-2.5pt]{
\draw[Aactivity] (0.5,0) -- (0,0) node[at end,above] {$\kappa$};
\draw[Aactivity] (0,0) -- (-0.5,0) node[at end,above] {};
\draw[substrate] (-130:0.5) -- (0,0);
\draw[substrate] (-50:0.5) -- (0,0);
}\\
\tikz[baseline=-2.5pt]{
\draw[Aactivity] (0.5,0) -- (0,0) node[at end,above] {$\chi$};
\draw[substrate] (0,0) -- (-0.5,0) node[at end,above] {};
\draw[substrate] (-130:0.5) -- (0,0);
\draw[substrate] (-50:0.5) -- (0,0);
}
&\qquad&
\tikz[baseline=-2.5pt]{
\draw[Aactivity] (0.5,0) -- (0,0) node[at end,above] {$-\lambda$};
\draw[substrate] (0,0) -- (-0.5,0) node[at end,above] {};
\draw[substrate] (-50:0.5) -- (0,0);
}.
\end{eqnarray}
Finally, the observables of the form of \Eref{observable_example} have the diagrammatic structure
\begin{equation} \elabel{diag_observable}
\tikz[baseline=-2.5pt]{
\begin{scope}[rotate=-30]
  \draw [decorate,decoration={brace,amplitude=5pt}] (-128:1.4cm) -- (-172:1.4) 
  %node[pos=0.5,left,xshift=-0.07cm,yshift=-0.14cm] {$n$};
  node[pos=0.5,left,xshift=-0.07cm] {$p$};
  \draw[substrate] (-130:0.3) -- (-130:1.3);
  \path [postaction={decorate,decoration={raise=0ex,text along path,
  text align={center},
  text={|\large|....}}}]
  (-155:1.2cm) arc (-155:-130:1.2cm);
  \draw[substrate] (-155:0.3) -- (-155:1.3);
  \draw[substrate] (-170:0.3) -- (-170:1.3);
\end{scope}
\draw[Aactivity] (0:0.3) -- (0:1.3);
\draw[thick,fill=white] (0,0) circle (0.3cm) 
node {};
} \ .
\end{equation}
Their scaling in time and finite-size can be extracted from the scaling of the vertex generating function, which is the standard object of field-theoretic renormalisation. In the next section we describe all possible infrared-relevant interactions.

\subsection{Relevant interactions} \slabel{Sec-RelevantInter}
Whether 
a particular interaction is allowed by the basic process introduces above is a matter of some topological constraints, 
which we will discuss in the first part of this section. Whether it is infrared-relevant is determined by its 
engineering dimension, which we discuss in the second part of this section. Combining topological and engineering constraints 
will then produce a finite number of interaction
vertices to consider. Constraints that avoid certain, otherwise relevant vertices 
from being generated are preserved under renormalisation.

The general proper vertex
\begin{equation} \elabel{general_proper_vertex}
\Gammaai{m}{n}{p}{q} = 
\tikz[baseline=-2.5pt]{
\begin{scope}[rotate=0]
  \draw [decorate,decoration={brace,amplitude=5pt}] (-128:1.4cm) -- (-172:1.4) node[pos=0.5,left,xshift=-0.07cm,yshift=-0.14cm] {$p$};
  \draw[substrate] (-130:0.3) -- (-130:1.3);
  \path [postaction={decorate,decoration={raise=0ex,text along path,
  text align={center},
  text={|\large|....}}}]
  (-155:1.2cm) arc (-155:-130:1.2cm);
  \draw[substrate] (-155:0.3) -- (-155:1.3);
  \draw[substrate] (-170:0.3) -- (-170:1.3);
\end{scope}
\begin{scope}[rotate=-60]
  \draw [decorate,decoration={brace,amplitude=5pt}] (-128:1.4cm) -- (-172:1.4) node[pos=0.5,left,xshift=-0.07cm,yshift=0.14cm] {$m$};
  \draw[Aactivity] (-130:0.3) -- (-130:1.3);
  \path [postaction={decorate,decoration={raise=0ex,text along path,
  text align={center},
  text={|\large|....}}}]
  (-155:1.2cm) arc (-155:-130:1.2cm);
  \draw[Aactivity] (-155:0.3) -- (-155:1.3);
  \draw[Aactivity] (-170:0.3) -- (-170:1.3);
\end{scope}
\begin{scope}[rotate=120]
  \draw [decorate,decoration={brace,amplitude=5pt}] (-128:1.4cm) -- (-172:1.4) node[pos=0.5,right,xshift=0.07cm,yshift=-0.14cm] {$q$};
  \draw[substrate] (-130:0.3) -- (-130:1.3);
  \path [postaction={decorate,decoration={raise=0ex,text along path,
  text align={center},
  text={|\large|....}}}]
  (-155:1.2cm) arc (-155:-130:1.2cm);
  \draw[substrate] (-155:0.3) -- (-155:1.3);
  \draw[substrate] (-170:0.3) -- (-170:1.3);
\end{scope}
\begin{scope}[rotate=180]
  \draw [decorate,decoration={brace,amplitude=5pt}] (-128:1.4cm) -- (-172:1.4) node[pos=0.5,right,xshift=0.07cm,yshift=0.14cm] {$n$};
  \draw[Aactivity] (-130:0.3) -- (-130:1.3);
  \path [postaction={decorate,decoration={raise=0ex,text along path,
  text align={center},
  text={|\large|....}}}]
  (-155:1.2cm) arc (-155:-130:1.2cm);
  \draw[Aactivity] (-155:0.3) -- (-155:1.3);
  \draw[Aactivity] (-170:0.3) -- (-170:1.3);
\end{scope}
\draw[thick,fill=white] (0,0) circle (0.3cm) 
node {};
}
\end{equation}
are the one-particle irreducible graphs of the amputated correlation function
\begin{eqnarray}
\Gai{m}{n}{p}{q}
\left(
r,D,\tau,s,\sigma,\lambda,\kappa,\chi,\xi;
\{\kvec_1,\ldots,\kvec_{m+n+p+q}; 
  \omega_1,\ldots,\omega_{m+n+p+q}\}
\right) \nonumber \\ 
=
\left\langle
\underbrace{\phi(\kvec_1,\omega_1)\ldots\phi(\kvec_m,\omega_m)}_{\textcolor{red}{m} \text{ terms}}
\underbrace{\psi\ldots\psi}_{\textcolor{olive}{p} \text{ terms}}
\underbrace{\phitilde\ldots\phitilde}_{\textcolor{red}{n} \text{ terms}}
\underbrace{\psitilde\ldots\psitilde}_{\textcolor{olive}{q} \text{ terms}}
\right\rangle &&
\end{eqnarray}
Denoting, where applicable, terms of higher order in non-linear couplings 
by 
\newcommand{\hot}{\text{h.o.t.}}
$\hot$, the bare couplings are the tree-level contributions to the 
proper vertices:
\begin{subequations}
\begin{equation}
\tau = \Gammaai{0}{1}{1}{0} + \hot 
\qquad
s = \Gammaai{2}{1}{0}{0}
\qquad
\lambda = \Gammaai{0}{1}{1}{1} + \hot
\qquad
\end{equation}
\begin{equation}
\sigma = \Gammaai{1}{1}{1}{0} + \hot
\qquad
\chi    = \Gammaai{0}{1}{2}{1} + \hot
\qquad
\kappa  = \Gammaai{1}{1}{1}{1} + \hot
\qquad
\xi     = \Gammaai{1}{1}{2}{1} + \hot
\end{equation}
\end{subequations}
Every proper vertex has a number of topological constraints, since any such term needs
to arise from the perturbative 
expansion of the action as a one-particle irreducible (connected, amputated) diagram 
made from the bare vertices available in the theory. By inspection, we found the following constraints, which we will use to determine all relevant, possible couplings below:
Firstly, all non-linear vertices in the field theory (all diagrams except the bare propagator of the tracer particles) have at least one straight leg coming in, $n\ge1$.
Secondly, all vertices have at least as many wavy legs coming out, as come in, $p\ge q$.
Thirdly, there are at least as many outgoing legs (wavy or straight), as there are incoming straight legs, 
$m+p\ge n$.

The engineering dimension of the general proper vertex can be determined from the considerations
at the beginning of \Sref{Sec_Dim_An_Couplings}, 
using the fact that each proper vertex may be seen as 
an effective coupling, which, after integration over real time and space,
gives rise to a dimensionless contribution to the action,
$\Ldim^d\Tdim\sbraket{\Gammaai{m}{n}{p}{q}\phitilde^m \psitilde^p \phi^n \psi^q}=1$, 
so that
\begin{equation}
\sbraket{\Gammaai{m}{n}{p}{q}} = 
\Ldim^{d(n+q-1)+2(m-n+2p-2q-1)}
\Adim^{p-q}
\Bdim^{m-n+p-q}
\Cdim^{n-m-2p+2q+1}.
\end{equation}
Demanding that (effective) transmutation $\tau$, branching $s$ 
and diffusion $D$ may remain relevant at any scale (which amounts
to a suitable continuum limit), we set the independent dimensions 
$\Adim$, $\Bdim$ and $\Cdim$, respectively,
to unity $\Adim=\Bdim=\Cdim=1$. The (marginally) 
infrared-relevant couplings are
those whose engineering dimension (in $\Ldim$) is non-positive. At the 
upper critical dimension
$d=d_c=4$, the inequality 
$d(n+q-1)+2(m-n+2p-2q-1)\le0$ gives
\begin{equation}\elabel{engineering_constraint}
m+n+2p \leq 3.
\end{equation}
The field theory needs to include all vertices 
$\Gammaai{m}{n}{p}{q}$ with
(non-negative) integers $m$, $n$, $p$ and $q$ that fulfill
\Eref{engineering_constraint} together with the topological 
constraints $n\ge1$, $p\ge q$ and $m+p\ge n$.
To find them, we distinguish two cases for \Eref{engineering_constraint}:
\begin{itemize}
\item $p=0 \overset{p\geq q}{\implies} q=0$, then $m+n \leq 3$. Under the topological constraint $m+p\geq n$ there are only two viable solutions: $m=n=1$, or $m=2$ and $n=1$, that correspond to 
\begin{equation}\label{Vert_relevant1}
\tikz[baseline=-2.5pt]{
\draw[Aactivity] (0.5,0) -- (-0.5,0);
}
\qquad \text{and} \qquad
\tikz[baseline=-2.5pt]{
\draw[Aactivity] (0.5,0) -- (0,0) node[at end,above] {$\ratebranch$};
\draw[Aactivity] (130:0.5) -- (0,0);
\draw[Aactivity] (-130:0.5) -- (0,0);
},
\end{equation}
the bare propagator for active walkers, and branching of active walkers, respectively. 

\item $p=1 \implies m+n \leq 1$. Only the propagator of the 
immobile particles allows for $n=0$. Otherwise,
$n\geq 1$ requires $m=0$. The constraint
$p\geq q$ 
leaves only $q=0$ and $q=1$. As a result, there are
three viable combinations: 
Firstly, $m=n=0$ and $q=1$, secondly,
$m=q=0$ and $n=1$, thirdly,
$m=0$ and $n=q=1$, which correspond to 
\begin{equation}\label{Vert_relevant2}
\tikz[baseline=-2.5pt]{
\draw[substrate] (0.5,0) -- (-0.5,0);
}
\,\text{,} \qquad
\tikz[baseline=-2.5pt]{
\draw[Aactivity] (0.5,0) -- (0,0) node[at end,above] {$\tau$};
\draw[substrate] (0,0) -- (-0.5,0) node[midway,above] {};
}
\qquad \text{and} \qquad
\tikz[baseline=-2.5pt]{
\draw[Aactivity] (0.5,0) -- (0,0) node[at end,above] {$-\lambda$};
\draw[substrate] (0,0) -- (-0.5,0) node[at end,above] {};
\draw[substrate] (-50:0.5) -- (0,0);
}, 
\end{equation}
the bare propagator of immobile tracer particles,
the transmutation vertex and 
hindrance of spawning, respectively.
\end{itemize}

Together with the propagators, the vertices in (\ref{Vert_relevant1}) and (\ref{Vert_relevant2}) represent all (marginally) relevant couplings at $d=d_c=4$, 
consisting of the (bilinear) transmutation, $\tau$, and 
the interaction vertices $s$ of branching and $-\lambda$ of 
suppression of spawning.

In the following we perform the renormalisation of the couplings $\tau$ and $-\lambda$.

\subsection{Renormalisation of the couplings}
\slabel{renorm}
As far as the observables in the present work are concerned, the only couplings to consider are $\tau$ and $\lambda$. Both are renormalised by the same set of loops
\begin{equation}\elabel{tau_renorm}
	\tau_R \corresponds
    \tikz[baseline=-2.5pt]{
\draw[substrate] (0,0) -- (-0.3,0) node [at start,above] {$\tau_R$};
\draw[Aactivity] (0.3,0) -- (0,0);
\fill (0,0) circle (3pt);
}
= \tau + 
		\tikz[baseline=-2.5pt]{
\draw[Aactivity] (0.4,0) arc (0:180:0.4);
\draw[Bsubstrate] (-0.4,0.0) arc (180:290:0.4);
\draw[Aactivity] (0.4,0.0) arc (0:-70:0.4);
\draw[Aactivity] (0.5,0) -- (0.4,0);
\begin{scope}[xshift=-0.4cm]
  \draw[substrate] (0,0) -- (-0.1,0);
\end{scope}
}
+
\tikz[baseline=-2.5pt]{
\draw[Aactivity] (0.4,0) arc (0:180:0.4);
\draw[Aactivity] (0.4,0.0) arc (0:-70:0.4);
\draw[Aactivity] (0.5,0) -- (0.4,0);
\draw[Aactivity] ($(0,0)+(237:0.39)$) -- ($(0,0)+(53:0.4)$);
\draw[Bsubstrate] (-0.4,0.0) arc (180:290:0.4);
\begin{scope}[xshift=-0.4cm]
  \draw[substrate] (0,0) -- (-0.1,0);
\end{scope}
}
+
\tikz[baseline=-2.5pt]{
\draw[Aactivity] (0.4,0) arc (0:180:0.4);
\draw[Aactivity] (0.4,0.0) arc (0:-70:0.4);
\draw[Aactivity] (0.5,0) -- (0.4,0);
\draw[Aactivity] ($(0,0)+(260:0.39)$) -- ($(0,0)+(30:0.4)$);
\draw[Aactivity] ($(0,0)+(214:0.38)$) -- ($(0,0)+(76:0.4)$);
\draw[Bsubstrate] (-0.4,0.0) arc (180:290:0.4);
\begin{scope}[xshift=-0.4cm]
  \draw[substrate] (0,0) -- (-0.1,0);
\end{scope}
}
+ \ldots 
\tikz[baseline=-2.5pt]{
\draw[Aactivity] ($(0,0)+(214:0.39)$) -- ($(0,0)+(30:0.4)$);
\draw[white,line width=3pt] ($(0,0)+(260:0.4)$) -- ($(0,0)+(76:0.4)$);
\draw[Aactivity] (0.4,0) arc (0:180:0.4);
\draw[Aactivity] (0.4,0.0) arc (0:-70:0.4);
\draw[Aactivity] (0.5,0) -- (0.4,0);
\draw[Aactivity] ($(0,0)+(260:0.40)$) -- ($(0,0)+(76:0.4)$);
\draw[Bsubstrate] (-0.4,0.0) arc (180:290:0.4);
\begin{scope}[xshift=-0.4cm]
  \draw[substrate] (0,0) -- (-0.1,0);
\end{scope}
}
+ \ldots
\end{equation}
and
\begin{equation}\elabel{lambda_renorm}
	-\lambda_R \corresponds
    \tikz[baseline=-2.5pt]{
\draw[Aactivity] (0.3,0) -- (0,0) node[at end,above] {$-\lambda_R$};
\draw[substrate] (0,0) -- (-0.3,0) node[at end,above] {};
\draw[substrate] (-45:0.3) -- (0,0);
\fill (0,0) circle (3pt);
} = -\lambda + 
		\tikz[baseline=-2.5pt]{
\draw[Aactivity] (0.4,0) arc (0:180:0.4);
\draw[Bsubstrate] (-0.4,0.0) arc (180:290:0.4);
\draw[Bsubstrate] (290:0.4) -- (290:0.5);
\draw[Aactivity] (0.4,0.0) arc (0:-70:0.4);
\draw[Aactivity] (0.5,0) -- (0.4,0);
\begin{scope}[xshift=-0.4cm]
  \draw[substrate] (0,0) -- (-0.1,0);
\end{scope}
}
+
\tikz[baseline=-2.5pt]{
\draw[Aactivity] (0.4,0) arc (0:180:0.4);
\draw[Aactivity] (0.4,0.0) arc (0:-70:0.4);
\draw[Aactivity] (0.5,0) -- (0.4,0);
\draw[Aactivity] ($(0,0)+(237:0.39)$) -- ($(0,0)+(53:0.4)$);
\draw[Bsubstrate] (-0.4,0.0) arc (180:290:0.4);
\draw[Bsubstrate] (290:0.4) -- (290:0.5);
\begin{scope}[xshift=-0.4cm]
  \draw[substrate] (0,0) -- (-0.1,0);
\end{scope}
}
+
\tikz[baseline=-2.5pt]{
\draw[Aactivity] (0.4,0) arc (0:180:0.4);
\draw[Aactivity] (0.4,0.0) arc (0:-70:0.4);
\draw[Aactivity] (0.5,0) -- (0.4,0);
\draw[Aactivity] ($(0,0)+(260:0.39)$) -- ($(0,0)+(30:0.4)$);
\draw[Aactivity] ($(0,0)+(214:0.38)$) -- ($(0,0)+(76:0.4)$);
\draw[Bsubstrate] (-0.4,0.0) arc (180:290:0.4);
\draw[Bsubstrate] (290:0.4) -- (290:0.5);
\begin{scope}[xshift=-0.4cm]
  \draw[substrate] (0,0) -- (-0.1,0);
\end{scope}
}
+ \ldots 
\tikz[baseline=-2.5pt]{
\draw[Aactivity] ($(0,0)+(214:0.39)$) -- ($(0,0)+(30:0.4)$);
\draw[white,line width=3pt] ($(0,0)+(260:0.4)$) -- ($(0,0)+(76:0.4)$);
\draw[Aactivity] (0.4,0) arc (0:180:0.4);
\draw[Aactivity] (0.4,0.0) arc (0:-70:0.4);
\draw[Aactivity] (0.5,0) -- (0.4,0);
\draw[Aactivity] ($(0,0)+(260:0.40)$) -- ($(0,0)+(76:0.4)$);
\draw[Bsubstrate] (-0.4,0.0) arc (180:290:0.4);
\draw[Bsubstrate] (290:0.4) -- (290:0.5);
\begin{scope}[xshift=-0.4cm]
  \draw[substrate] (0,0) -- (-0.1,0);
\end{scope}
}
+ \ldots
\end{equation}
where all diagrams are amputated. The subscript $R$ indicates a renormalised quantity.
Only the non-crossing loop diagrams, such as the first three in \Erefs{tau_renorm} and
\eref{lambda_renorm}, are easily calculated (see \suppsref{integrals} for details).
Of the diagrams in \Erefs{tau_renorm} and \eref{lambda_renorm}, the non-crossing
ones are summed over by virtue of field-theoretic renormalisation.
The last diagram in both \Eref{tau_renorm} and \Eref{lambda_renorm}, on the other hand,
require further 
explicit calculation and subsequent summation. The same applies to an infinite number
of further crossing diagrams. And yet, because of the Ward-identity
\begin{equation}
	\ddXX{\tau_R}{\tau}=
	\ddXX{\lambda_R}{\lambda}
\end{equation}
all exponents can be determined without calculating any of the diagrams explicitly.

As usual in perturbative field theory \citep{STaeuber:2014,SLeBellac:1991}, the governing non-linearity, here $\lambda$, becomes spatially dimensionless by multiplying it by $\mu^{-\epsilon}$,
where $\mu$ is an arbitrary inverse length scale.
In fact, \emph{any} dimensionless coupling involving
$\lambda$, $\tau$, $s$, $D$ and $\mu$ is proportional
to a power of $\lambda s D^{-2} \mu^{-\epsilon}$. Introducing
$g=\lambda s \UC \mu^{-\epsilon} D^{-2} \Gamma(\epsilon/2)$
with suitable numerical factor $\UC$, both
couplings $\lambda$ and $\tau$ renormalise identically
\begin{equation}
	\tau_R=\tau Z(g) \qquad \text{ and } \qquad \lambda_R = \lambda Z(g)
\end{equation}
with $Z(g)$ governing the renormalisation of both $\lambda$
and $\tau$.
To one loop and with suitable $\UC$, the $Z$-factor becomes
$Z(g)=1-g$, see \Erefs{tau_renorm} and \eref{lambda_renorm}, and \sref{integrals}. However, there is no need to
determine the precise dependence of $Z$ on $g$ as far as
scaling is concerned. It suffices to know that the
renormalised, dimensionless 
\begin{align}
g_R&=\lambda_R s \UC \mu^{-\epsilon} D^{-2} \Gamma(\epsilon/2) \\
&= Z \lambda s \UC \mu^{-\epsilon} D^{-2} \Gamma(\epsilon/2)
\end{align}
has $\beta$-function
\begin{equation}
    \beta_g=\ddXX{g_R}{\ln \mu}
= - \epsilon g_R + g_R \frac{\plaind \ln Z}{\plaind \ln \mu}
\end{equation}
which implies $\plaind \ln Z/\plaind \ln \mu=\epsilon$
at the root $\beta_g(g=g^*)=0$, irrespective of 
$\UC$ and therefore irrespective of the presence or absence of the crossing diagrams. It follows that $Z\sim \mu^\epsilon$ in $d\le4$ and therefore the effective transmutation rate is
$\tau_{\eff}\sim \tau Z \sim\mu^\epsilon$.
In the limit of $t\to\infty$, for systems of linear size $L$,
the characteristic scale is $\mu\sim L^{-1}$ and
thus $\tau_{\eff}\sim L^{-\epsilon}$. With open boundary conditions, the branching walkers visit 
$\sim L^2$ sites during the course of their lifetimes,
leaving behind $\sim \tau_{\eff} L^2 \sim L^{2-\epsilon}$
immobile tracer particles in dimensions greater than $2$,
so that
$\ave{\dsv}(t,L)\sim L^{d-2}$. 
This average is bounded from below by a constant, as at least one site is always visited, so that $\ave{\dsv}(t,L)$
approaches a constant
below $2$ dimensions.
As for the time-dependence, 
the characteristic inverse scale $\mu$ is proportional
to $t^{-1/2}$
because the dynamical exponent $z=2$ in $\mu \sim t^{-1/z}$ remains unchanged.
It follows that $\ave{\dsv}(t,L)\sim t^{(d-2)/2}$.

In the following section, the mean $\ave{\dsv}(t,L)$ and
higher moments are calculated in greater detail.

\begin{comment}
\hrule
To see this, we introduce a dimensionless coupling 
$g=\lambda s \UC \mu^{-\epsilon} D^{-2} \Gamma(\epsilon/2)$ with suitable
numerical
factor
$\UC$
and arbitrary inverse length scale $\mu$. 
Both couplings therefore renormalise identically,
\begin{equation}
	\tau_R=\tau Z \qquad \text{ and } \qquad \lambda_R = \lambda Z
\end{equation}
with $Z=1-g$
governing the renormalisation of both renormalised $\tau$ and $\lambda$.
The $\beta$-function of $g$,
\[
\beta_g=\ddXX{g}{\ln \mu}
\]
in $d<4$
always produces a root $g^*$ such that $\gamma_{\tau}=\ddXX{\ln Z}{\ln \mu}=\epsilon = 4-d$
when $g=g^*$, irrespective of $\UC$ and therefore irrespective of the presence or absence of the crossing diagrams. 
Because the $Z$-factor for $\lambda$ is identical to that of $\tau$, the latter scales in the inverse scale $\mu$ like 
$\mu^{\epsilon}$. As the dynamical exponent $z=2$ remains unchanged, it follows that in the long time $t\rightarrow \infty$, and large system size $L\rightarrow \infty$ limits, the volume visited 
$\ave{\dsv}(t,L)$ by a walker
by time $t$ scales like $t^{(d-2)/2}$ in dimensions $d<4$. It remains finite in dimensions $d<2$, as discussed in the main text.

\end{comment}

\subsection{Calculating scaling of higher-order correlation functions} \slabel{Callan_Sym}
The scaling of higher-order correlation functions is derived, within the field theory, from the solution of the Callan–Symanzik equation \citep{STaeuber:2014} for the general proper vertex \Eref{general_proper_vertex}, from which the scaling of the moments of the total number of distinct sites visited follow, \Eref{moment_result}.
From dimensional analysis (\sref{Sec_Dim_An_Couplings}), and by introducing a bare scale $\mu_0$, related to $\mu$ by $\mu = \mu_0 \ell$, the general proper vertex, \Eref{general_proper_vertex}, then satisfies
\begin{multline}
\Gammaai{m}{n}{p}{q}(r,D,\tau,\ratebranch,\sigma,\lambda,\kappa,\chi,\xi;\{\kvec;\omega\}) \\
= \ell^{-d(n+q-1)-2(m-n+2p-2q-1)+(p-q)\gamma_\tau} \Gammaai{m}{n}{p}{q}\left(\frac{r}{\ell^2},D,\tau,\ratebranch,\sigma,\lambda,\kappa,\chi,\xi;\Big\{ \frac{\kvec}{\ell};\frac{\omega}{\ell^2}\Big\} \right),
\end{multline}
asymptotically in small $\ell$ and provided that $r$ is close enough to the critical point, $r_c=0$. For the transmutation vertex, where $p=n=1$ and $q=m=0$, we find
\begin{equation} \elabel{gamma_first_mom}
\Gammaai{0}{1}{1}{0}(r,D,\tau,\ratebranch,\sigma,\lambda,\kappa,\chi,\xi;\{\kvec;\omega\}) = \ell^{\gamma_\tau} \Gammaai{0}{1}{1}{0}\left(\frac{r}{\ell^2},D,\tau,\ratebranch,\sigma,\lambda,\kappa,\chi,\xi;\Big\{ \frac{\kvec}{\ell};\frac{\omega}{\ell^2} \Big\} \right),
\end{equation}
with $\gamma_\tau = \varepsilon = 4-d$. Generally, for observables of the form \Eref{diag_observable}, where $n=1$ and $q=m=0$ we have
\begin{equation} \elabel{gamma_high_mom}
\Gammaai{0}{1}{p}{0}(r,D,\tau,\ratebranch,\sigma,\lambda,\kappa,\chi,\xi;\{\kvec;\omega\}) = \ell^{4(1-p)+p\gamma_\tau} \Gammaai{0}{1}{p}{0}\left(\frac{r}{\ell^2},D,\tau,\ratebranch,\sigma,\lambda,\kappa,\chi,\xi;\Big\{ \frac{\kvec}{\ell};\frac{\omega}{\ell^2} \Big\} \right).
\end{equation}
The scaling of the first moment of the number of distinct sites visited, $\ave{\dsv(t)}$, as function of time, t, can be obtained by analysing the scaling of  
\begin{eqnarray}
\ave{\dsv(t)} &=& \int\plaind^d\xvec\ave{\psi(\xvec,t)\phitilde(\xvec_0,0)} \\
& \corresponds & \left. \int \dbar\omega \dbar\omega_0 \exp{-\imag\omega t}
\tikz[baseline=-2.5pt]{
\draw[substrate] (0,0) -- (-0.5,0) node [at start,above] {};
\draw[Aactivity] (0.5,0) -- (0,0);
\fill (0,0) circle (3pt);
} \right|_{\kvec=0} \\
&=& \int \dbar\omega \exp{-\imag\omega t}\frac{1}{-\imag\omega+\rateextB}\Gammaai{0}{1}{1}{0}\frac{1}{-\imag\omega+r}.
\elabel{int_omega}
\end{eqnarray}
According to \Eref{gamma_first_mom}, $\Gammaai{0}{1}{1}{0}$ scales like
\begin{equation} \elabel{gamma_first_mom_scale}
\Gammaai{0}{1}{1}{0}(L^{-2},D,\tau,\ratebranch,\sigma,\lambda,\kappa,\chi,\xi;\{\kvec;\omega\}) = L^{-\gamma_\tau} \Gammaai{0}{1}{1}{0}\left(1,D,\tau,\ratebranch,\sigma,\lambda,\kappa,\chi,\xi;\Big\{\kvec L;\omega L^2 \Big\} \right),
\end{equation} 
if we identify $r \sim L^{-2}$ and $\ell \sim L^{-1}$, which means that the effective transmutation rate scales like $L^{-\varepsilon}$ in large linear system size $L$, as $\gamma_\tau=\varepsilon=4-d$. In long time $t$, the integral over $\omega$ in \Eref{int_omega} has the effect of evaluating $\Gammaai{0}{1}{1}{0}\frac{1}{\imag\omega+r}$ at $\omega=0$, 
because 
\begin{equation}
    \lim_{t\to\infty}\lim_{\epsilon'\to0}
    \int_{-\infty}^{\infty}
    \exp{-\imag \omega (t-t_0)} 
    \frac{1}{-\imag \omega + \epsilon'} f(\omega) =
    f(0)
\end{equation}
provided $f(\omega)$ has no pole at $0$.

It follows that
\begin{equation}
\lim_{t\to \infty} \ave{a(t)} \propto L^{2-\epsilon}.
\end{equation}
For higher moments, on the basis of \Eref{gamma_high_mom} we find
\begin{equation}
\lim_{t\to \infty} \ave{a^p(t)} \propto L^2L^{pd-4}\Gammaai{0}{1}{p}{0}(1,D,\tau,\ratebranch,\sigma,\lambda,\kappa,\chi,\xi;\{0,0\}).
\end{equation}
We thus recover the finite-size scaling results 
\Erefs{moment_result_in_L} and \eref{moment_result_above_dc_in_L}
of \Sref{Results_regular} for the $p$-th moment of the
volume explored by a branching random walk
\begin{equation}
\lim_{t\to \infty} \ave{a^p(t)} \propto
\begin{cases}
L^{dp-2} & \text{if $\varepsilon > 0$} \\
L^{4p-2} & \text{if $\varepsilon < 0$}
\end{cases}
\end{equation}
where $\varepsilon>0$ and $\varepsilon<0$ separate regions below and above the upper critical dimension, $d_c=4$, respectively.
The dimensionality of the embedding space enters only below the upper critical dimension. 
Above the upper critical dimension, fluctuations and interactions become asymptotically irrelevant
and the process can be considered as free.

The above analysis is easily extended to scaling in time, using $t \propto \mu^{-z}$ with $z=2$ as the relevant scale, thereby
reproducing \Erefs{moment_result_in_t} and \eref{moment_result_above_dc_in_t}.

\section{Loop integrals}\slabel{integrals}
The non-crossing diagrams, such as the first three in \Erefs{tau_renorm} and
\eref{lambda_renorm}, are calculated through the integral

\begin{equation}\elabel{main_loop_tau}
	I_\tau
=\tikz[baseline=-2.5pt]{
\draw[Aactivity] (0.4,0) arc (0:180:0.4) node [at start,above,xshift=0.15cm] {$s$} node [at end,above,xshift=-0.25cm] {$-\lambda$};
\draw[Bsubstrate] (-0.4,0.0) arc (180:270:0.4);
\draw[Aactivity] (0,-0.4) arc (270:360:0.4) node [at start,below] {$\tau$} node [at start,above] {};
\draw[Aactivity] (0.5,0) -- (0.4,0);
\begin{scope}[xshift=-0.4cm]
  \draw[substrate] (0,0) -- (-0.1,0);
\end{scope}
} 
=
\int\ddintbar{k}\dintbar{\omega}
\frac{\tau}{-\imag\omega+\rateextB}
\frac{1}{\omega^2+(Dk^2+\mass)^2}
=
\tau\half \frac{r^{-\varepsilon/2}}{(4\pi D)^{d/2}}\Gamma(\varepsilon/2)
\ ,
\end{equation}
and (essentially identical)
\begin{equation}\elabel{main_loop_lambda}
	I_{-\lambda}
=\tikz[baseline=-2.5pt]{
\draw[Aactivity] (0.4,0) arc (0:180:0.4) node [at start,above,xshift=0.15cm] {$s$} node [at end,above,xshift=-0.25cm] {$-\lambda$};
\draw[Bsubstrate] (-0.4,0.0) arc (180:270:0.4);
\draw[Aactivity] (0,-0.4) arc (270:360:0.4) node [at start,below] {$-\lambda$} node [at start,above] {};
\draw[Aactivity] (0.5,0) -- (0.4,0);
\begin{scope}[xshift=-0.4cm]
  \draw[substrate] (0,0) -- (-0.1,0);
\end{scope}
\begin{scope}[yshift=-0.4cm]
  \draw[substrate] (0,0) -- (0.0,-0.1);
\end{scope}

} 
=
\int\ddintbar{k}\dintbar{\omega}
\frac{-\lambda}{-\imag\omega+\rateextB}
\frac{1}{\omega^2+(Dk^2+\mass)^2}
=
-\lambda\half \frac{r^{-\varepsilon/2}}{(4\pi D)^{d/2}}\Gamma(\varepsilon/2)
\ ,
\end{equation}

where the lower part of the loop carries the coupling $\tau$ in case of contributing to $\tau$ 
or the coupling $-\lambda$ and an incoming wavy leg in case of contributing
to $\lambda$. The integration measure is $\ddintbar{k}\dintbar{\omega}=\ddint{k}\dint{\omega}/(2\pi)^{d+1}$.

\section{Generalisation to $k$ offspring}\slabel{k_offspring}
In this section we extend the field-theoretic results presented 
above to the case where the offspring number is a random number and show that it lies in the same universality class as binary branching \citep{Saldous1993continuum,Sle2005random}.
Instead of two distinct processes for branching into two active
walkers (with rate $\ratebranch$ above) and getting extinguished (with rate $\rateextA$ above)
we consider the latter as branching into $k=0$ walkers and generalise the
former to branching into any number $k$ of walkers. 
Each of these processes may occur with rate $\sigma_k$, which can 
always be written as $\sigma_k=\sigma p_k$ with $p_k$ the normalised probability
for branching into $k$ walkers and $\sigma$ the rate with which any such
processes take place.

The two contributions 
$\Prob_{\ratebranch}$, \Eref{ME_branching}, 
and 
$\Prob_{\rateextA}$, \Eref{ME_extA}, 
are thus subsumed
and generalised by
\begin{equation}
\dot{\Prob}_{c}(\{n\},\{m\};t) = 
\sigma
\sum_{k=0}^{\infty}\sum_{\xvec}
 p_k\Big(  (n_{\xvec}-k+1) \Prob(\{\ldots,n_{\xvec}-k+1,\ldots\},\{m\};t)
                - n_{\xvec} \Prob(\{n\},\{m\};t) \Big),
\end{equation}
which allows for $p_1$, but the process
of branching into a single particle has no bearing on the master equation.

In the field theory, the mass of the bare propagator for active walkers becomes \citep{SGarcia-MillanETAL:2018}
\begin{equation}
\mass = - \sigma \sum_{k=0}^{\infty} p_k(k-1) \\
= \sigma (1-\bar{k}),
\end{equation}
where $\bar{k} = \sum_{k=0}^{\infty} p_k k$ is the average offspring number,
which again, defines a subcritical ($\mass >0$), a critical ($\mass = 0$), and a supercritical ($\mass <0$) regime.

In the case of generalised branching,
the non-linear part of the action 
contains
contributions of the form $\phitilde^k \phi$ for all $k\ge2$ as soon as
there is any $k\ge2$ with $p_k>0$
\cite{SGarcia-MillanETAL:2018}.
Terms with $k>2$, however, turn out to be infrared irrelevant,
as their couplings have dimension $\Bdim^{k-1}\Cdim^{2-k}\Ldim^{2(k-2)}$.
The field theoretic results above for binary branching 
therefore govern also branching 
processes with generalised offspring distribution.

\section{Extension to general graphs}
\slabel{gen_graph}
In this section we provide further details about the extension of our results to general graphs.
The loops integrated over in \Erefs{main_loop_tau} and \eref{main_loop_lambda} are
in fact integrals over
the spectrum of the Laplacian accounting for the diffusion on the graph considered.
Generalising to arbitrary graphs, the Laplacian is to be replaced by a lattice-Laplacian
and the integral in \Erefs{main_loop_tau} and \eref{main_loop_lambda} by a suitable sum or, equivalently, an
integral with
suitable spectral density. 
In fact, the $d$-dimensional integral in \Erefs{main_loop_tau} and \eref{main_loop_lambda} can be seen as an
integral over all distinct eigenvalues $\kvec^2$ of the Laplacian entering
with weight $w(k)\dint{k}=S_d k^{d-1} \dint{k}$ with $S_d=2\pi^{d/2}/\Gamma(d/2)$.
On regular lattices, their Hausdorff dimension $d$ coincides with the spectral
dimension $d_s$ characteristing, in particular, the small $k$ asymptote of
$w(k)\sim k^{d_s-1}$. 
Replacing $\int \ddint{k}$ by $\int \dint{k} w(k)$ suggests that the results derived above remain valid by replacing $d$ by $d_s$, in order to recover the scaling of the various observables in arbitrary graphs with spectral dimension $d_s$. 
The replacement $d\to d_s$ hinges crucially on the fact that $d_s$ characterises the
scaling of the spectral density of the Laplacian. If this operator itself renormalises, then 
a different spectral density may be needed. 
In other words, $d_s$ may not be 
the correct dimension if the Laplacian renormalises, \ie if the anomalous
dimension does not vanish, $\eta\ne0$ \citep{Sburioni2005}.
This argument relies on the assumption that vertices such as \Eref{general_proper_vertex} preserve momentum, 
that is integrals of the form
\begin{equation}
I_n(\kvec_1,\kvec_2,\ldots,\kvec_n)=
\int \ddint{x} 
u_{\kvec_1}(x)
u_{\kvec_2}(x)
\ldots
u_{\kvec_n}(x)
\end{equation}
over eigenfunctions $u_{\kvec}(x)$ 
of the Laplacian
with eigenvalue $\kvec\cdot\kvec$ vanish for off-diagonal terms, i.e. whenever $\kvec_1+\kvec_2+\ldots+\kvec_n\ne0$. 
This condition can be further relaxed by demanding merely that off-diagonal terms
are sub-leading as observed in the presence of boundaries \cite{SDiehlSchmidt:2011,SNekovarPruessner:2016}.

Considering only graphs which are translationally invariant
such that the indices $\jvec_m$ of the $q$ 
neighbours $m=1,\ldots,q$ of any node $\ivec$ can be determined by adding the same
set of translational lattice vectors, $\dvec_1,\ldots,\dvec_q$, such that
$\jvec_m = \ivec + \dvec_m$,
it is easy to show that the Laplacian has exponential eigenfunctions and any of their 
products
are an eigenfunction as well, so that 
$I_n(\kvec_1,\kvec_2,\ldots,\kvec_n)=I_2(\kvec_1,\kvec_2+\ldots+\kvec_n)$,
which vanishes by orthogonality for any
$\kvec_1+\kvec_2+\ldots+\kvec_n\ne\nullvec$, i.e. the assumption of momentum conservation mentioned above is fulfilled.

\section{Numerics for the scaling of moments} \slabel{fitting}
The scaling of the moments $\ave{\dsv^p}(t,L)$ for $p=1,2,3,\ldots ,5$, as function of time $t$ in the limit $L\rightarrow\infty$, and as function of the system size $L$ in the limit $t\rightarrow\infty$ were
obtained from numerical Monte Carlo simulations and fitted against a power-law 
\begin{equation} \elabel{fitting1}
f(x)=Ax^B
\end{equation}
and a power-law with corrections of the form
\begin{equation} \elabel{fitting2}
g(x)=Ax^B + Cx^{B-1/2}.
\end{equation}
The fitting parameter $B$ in \Erefs{fitting1} and \eref{fitting2} provides the estimates of the exponents that characterise the scaling of the moments in time $t$ and sistem size $L$ (or $N$, see main text), by fitting the numerical estimates against $f(x)$ and $g(x)$, with $x$ replaced by $t$ and $L$, respectively. At large times the moments display plateauing due to finite-size effects. 

For the scaling in system size $L$, we fitted the data for the latest time point available against \Eref{fitting1} and used the estimates of $A$ and $B$ as the initial values for a fit against \Eref{fitting2}, which gave the final estimates of the finite-size scaling exponents.

For the scaling in time $t$, we fitted data for the largest system, of size $L=L_{\text{max}}$. The fitting range in $t$ for each moment was determined systematically as follows. To remove the time-point affected by the finite-size effects, we defined the upper bound of the fitting range as the time $t^{\text{up}}$ for which the lowest moment displaying algebraic divergence ($p=p_{\text{low}}$) reached a value of half the maximum value in the plateau, {\it i.e.} $ \ave{\dsv^{p_{\text{low}}}} (t^{\text{up}},L_{\text{max}}) = \max\limits_t\left(\ave{\dsv^{p_{\text{low}}}} (t,L_{\text{max}})\right)/2$. For the preferential attachment network the plateau was observed to occur at an earlier time point than $t^{\text{up}}$, probably due to the high connectivity of the networks, so we set the upper bound to $t^{\text{up}}_{pa} = (1/5)\max(\ave{\dsv^k})$, in this case.

To find the lower bound $t_{\text{low}}$ of the fitting range in $t$ we fitted both equations, \eref{fitting1} and \eref{fitting2}, to the data for $L_{\text{max}}$. We define $\hat{f}_{[t^*,t^{\text{up}}]}(t)$ and $\sigma^{\hat{f}}_{[t^*,t^{\text{up}}]}(t)$ as the values and errors, respectively, of fitting \Eref{fitting1} to the data in the range $t\in [t^*,t^{\text{up}}]$, and $\hat{g}_{[t^*,t^{\text{up}}]}(t)$ and $\sigma^{\hat{g}}_{[t^*,t^{\text{up}}]}(t)$ as the values and errors, respectively, of fitting \Eref{fitting2} to the same data set and range. Further, we define $N_{[t^*,t^{\text{up}}]}$ as the number of data points within the fitting interval $[t^*,t^{\text{up}}]$. The lower bound for the time range $t_{\text{low}}$ is the earliest time at which both fitting models \eref{fitting1} and \eref{fitting2} agree within errors, that is

\begin{equation}
t_{\text{low}} = \min\Big\{t^*:\abs{\hat{f}_{[t^*,t^{\text{up}}]}(t^*)-\hat{g}_{[t^*,t^{\text{up}}]}(t^*)} \leq \sqrt{N_{[t^*,t^{\text{up}}]}}\max\left(\sigma^f_{[t^*,t^{\text{up}}]}(t^*),\sigma^g_{[t^*,t^{\text{up}}]}(t^*)\right)\Big\}.
\end{equation}
Where we account for correlations between estimates of moments by rescaling the error by the square root of the number of data points in the fitting range, $N_{[t^*,t^{\text{up}}]}$. The exponents characterising the time depenence of the moments are determined by fitting the data in the range $[t^*,t^{\text{up}}]$ against \Eref{fitting2}.

The fitting of the power laws, \Erefs{fitting1} and \eref{fitting2}, was done by means of the Levenberg-Marquardt algorithm \citep{SPressETAL:1992}. %All fittings presented here correspond to a goodness-of-fit greater than $1$. Fittings with lower values are indicated by an asterisk. 
In \suppTref{table_scaling_time} and \supptre{table_scaling_syst} we report the numerical results for the asymptotic scaling in time, $\ave{\dsv^p}(t)\sim t^{\alpha_p}$, and in system size, $\ave{\dsv^p}(t)\sim L^{\beta_p}$, provided these observables display an algebraic divergence.

\begin{table}[h!]
	\caption*{{\bf Scaling of visited sites in time}}
	\begin{tabular}{c||l|l||l|l||l|l||l|l||l|l||l|l||l|l||} 
		\multirow{2}{*}{exponent} & \multicolumn{2}{c||}{d=1} & \multicolumn{2}{c||}{d=2} & \multicolumn{2}{c||}{d=3} & \multicolumn{2}{c||}{d=5} & \multicolumn{2}{c||}{S.C.} & \multicolumn{2}{c||}{R.T.} & \multicolumn{2}{c||}{P.A.} \\
        \cline{2-15}
        & num & theo & num & theo & num & theo & num & theo & num & theo & num & theo & num & theo \\
		  \hline
		   $\alpha_1$ &  & & & & 0.47(2)  & 1/2 & 1.0(2) & 1 & & & & & 1.0(1) & 1 \\
           		   \hline
		   $\alpha_2$ &  & & 0.98(3) & 1 & 2.0(1) & 2 & 2.9(3) & 3 & 0.81(5) & 0.86 & 0.35(7) & 1/3 & 2.8(2) & 3\\
		    \hline
            $\alpha_3$ & 0.48(4) & 1/2 & 2.0(1) & 2 & 3.5(1) & 7/2 & 4.8(4) & 5 & 1.71(7) & 1.79 & 0.9(1) & 1 &  4.8(4) & 5\\
		     \hline
		   $\alpha_4$ & 1.0(1) & 1 & 2.9(1) & 3 & 5.0(2) & 5 & 6.7(7) & 7 & 2.62(10) & 2.72 & 1.6(2) & 5/3 &  6.6(5) & 7 \\
		     \hline
		   $\alpha_5$ & 1.5(1) & 3/2 & 3.9(1) & 4 &  6.4(2) & 13/2 & 9(1) & 9 & 3.54(14) & 3.66 & 2.2(4) & 7/3 & 8.5(9) & 9\\
		     \hline
             \hline
		   mean gap & 0.5(1) & 1/2 & 1.0(1) & 1 & 1.5(2) & 3/2 & 2.0(5) & 2 & 0.91(9) & 0.93 & 0.6(2) &2/3& 1.9(4) & 2 \\
		   \hline
           % Note: for PA we used a threshold 0.2*max(<a>) not 0.5 as in all others, this gives a range [[335.371, 1164.303]
	\end{tabular}
	\caption{Scaling in time, $\ave{\dsv^p}(t)\sim t^{\alpha_p}$, of the $p$-th moment of the number of distinct sites visited for regular lattices of integer dimension, $d$, as indicated, and for the Sierpinski carpet (S.C., $d_s\approx 1.86$), the random tree (R.T., $d_s=4/3$), and preferential attachment (P.A.,  $d_s > 4$) networks. The columns marked {\it num} shows the numerical results, the columns marked {\it theo} show theoretical results according to \Erefs{moment_result_in_t} with $d$ replaced by the spectral dimension $d_s$ where applicable. The row marked {\it mean gap} show the average gap-exponent, $(1/p)\sum_{i=1}^p(\alpha_{i+1}-\alpha_i)$,  for the corresponding lattice.}
    \tlabel{table_scaling_time}
\end{table}

\begin{table}[h!]
	\caption*{{\bf Scaling of visited sites by a BRW as function of the system size}}
	\begin{tabular}{c||l|l||l|l||l|l||l|l||l|l||l|l||l|l||} 
		\multirow{2}{*}{exponent} & \multicolumn{2}{c||}{d=1} & \multicolumn{2}{c||}{d=2} & \multicolumn{2}{c||}{d=3} & \multicolumn{2}{c||}{d=5} & \multicolumn{2}{c||}{S.C.} & \multicolumn{2}{c||}{R.T.} & \multicolumn{2}{c||}{P.A.}  \\
        \cline{2-15}
        & num & theo & num & theo & num & theo & num & theo & num & theo & num & theo & num & theo \\
		  \hline
		   $\beta_1$ & & & & & 0.97(4) & 1 & 1.9(2)* & 2 & & & & & 0.49(1) & 1/2\\
           		   \hline
		   $\beta_2$ & & & 2.1(2) & 2 & 3.9(1)* & 4 & 5.7(5)* & 6 & 1.9(1) & 1.72  & 0.58(6) & 1/2 & 1.49(1) & 3/2 \\
		    \hline
           $\beta_3$ & 0.96(4) & 1 & 4.2(3) & 4 & 6.8(2)* & 7 & 10(1) & 10 & 3.8(2) & 3.59  & 1.6(1) & 3/2 & 2.49(2) & 5/2 \\
		     \hline
		   $\beta_4$ & 1.93(7) & 2 & 6.1(3) & 6 & 9.8(3)* & 10 & 14(2) & 14 & 5.7(3) & 5.45 & 2.6(2) & 5/2 & 3.49(2) & 7/2 \\
		     \hline
		   $\beta_5$ & 2.93(8) & 3 & 8.1(4) & 8 & 12.7(4)* & 13 & 17(3) & 18 & 7.5(4) & 7.31  & 3.7(2) & 7/2 &4.49(2)& 9/2\\
		     \hline
             \hline
		   mean gap & 0.9(1) & 1 & 2.0(3) & 2 & 2.9(2) & 3 & 4(1) & 4 & 1.9(3) & 1.86 & 1.0(1) & 1 & 1.00(2) & 1 \\
           %\hline
		   \hline
           fit range & \multicolumn{2}{c||}{[255, 4095]} & \multicolumn{2}{c||}{[15, 127]}& \multicolumn{2}{c||}{[7, 127]} & \multicolumn{2}{c||}{[7, 31]} & \multicolumn{2}{c||}{[9, 243]} & \multicolumn{2}{c||}{[$2^6-1$, $2^{12}-1$]} & \multicolumn{2}{c||}{[$2^{14}-1$, $2^{19}-1$]}\\
	\end{tabular}
	\caption{Scaling, $\ave{\dsv^p}(t)\sim L^{\beta_p}$, of the $p$-th moment of the number of distinct sites visited as function of the system size $L$, for regular lattices of integer dimension $d$ as indicated and for the Sierpinski carpet (S.C., $d_s\approx 1.86$). The columns marked {\it num} show the numerical results, the columns marked {\it theo} show theoretical results according to \Eref{moment_result_in_L}, for regular lattices, and S.C. (with $d$ replaced by the spectral dimension $d_s$), and according to \Eref{moment_result_networks} for random tree (R.T.) and preferential attachment (P.A.) The row marked {\it mean gap} shows the average gap-exponent, $(1/p)\sum_{i=1}^p(\beta_{i+1}-\beta_i)$,  for the corresponding lattice. %The row maked {\it fit range} show the range of linear system sizes, $L$, used in the fitting of the exponents. 
 *Goodness of fit $<$ 0.05.}
    \tlabel{table_scaling_syst}
\end{table}

\clearpage

\section*{Supplementary Figures}

\begin{figure}[h!]
\includegraphics[width=0.45\textwidth]{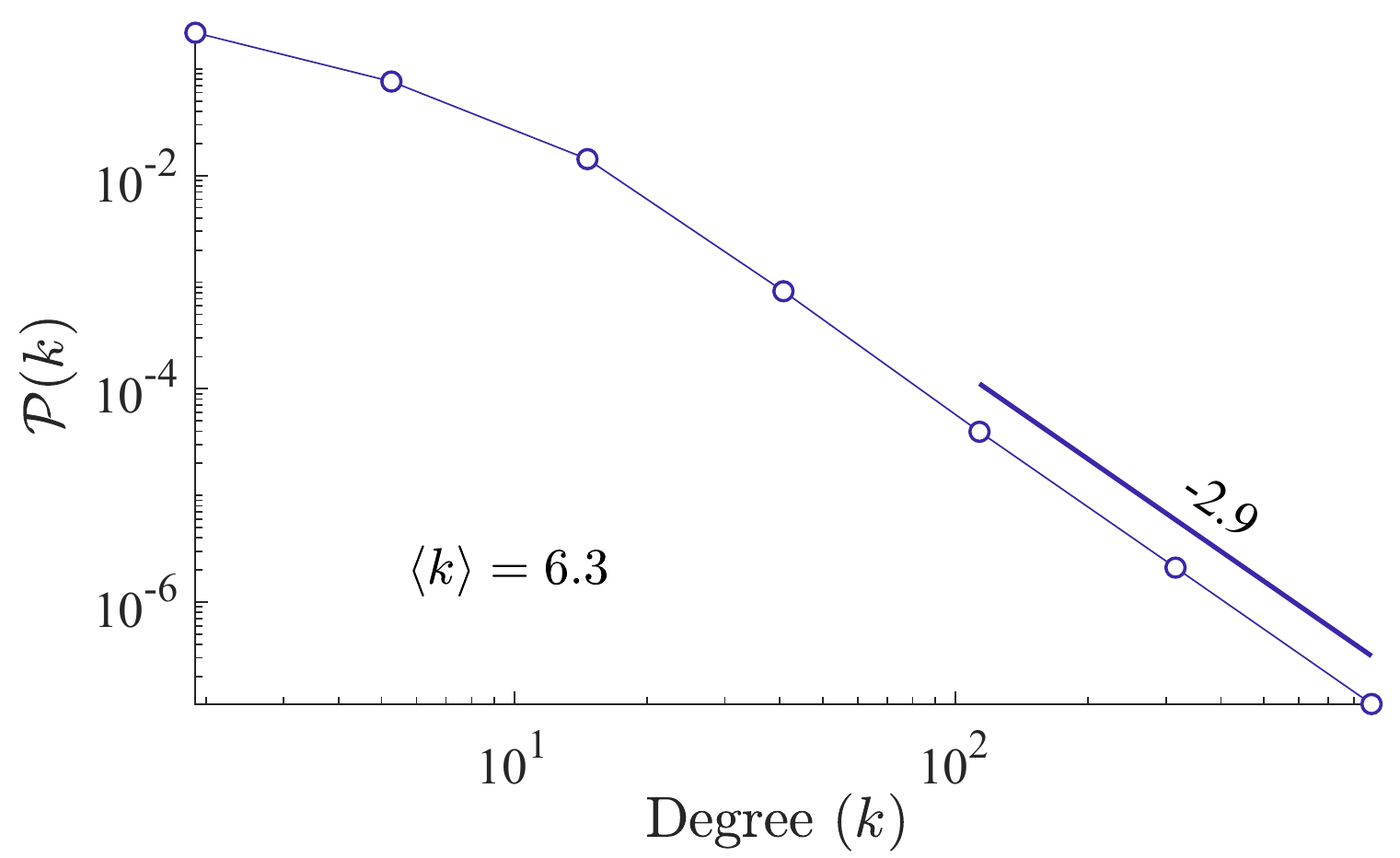}
\caption{{\bf Degree distribution of the preferential attachment networks}  used for the simulations presented in \sref{sec_gen_graphs}}
\flabel{FS2_degree_dist}
\end{figure}


\begin{thebibliography}{10}
\expandafter\ifx\csname url\endcsname\relax
  \def\url#1{\texttt{#1}}\fi
\expandafter\ifx\csname urlprefix\endcsname\relax\def\urlprefix{URL }\fi
\providecommand{\bibinfo}[2]{#2}
\providecommand{\eprint}[2][]{\url{#2}}

\bibitem{EubankETAL:2004}
\bibinfo{author}{Eubank, S.} \emph{et~al.}
\newblock \bibinfo{title}{Modelling disease outbreaks in realistic urban social
  networks}.
\newblock \emph{\bibinfo{journal}{Nature}} \textbf{\bibinfo{volume}{429}},
  \bibinfo{pages}{180--184} (\bibinfo{year}{2004}).

\bibitem{Pastor-SatorrasETAL:2015}
\bibinfo{author}{Pastor-Satorras, R.}, \bibinfo{author}{Castellano, C.},
  \bibinfo{author}{Van~Mieghem, P.} \& \bibinfo{author}{Vespignani, A.}
\newblock \bibinfo{title}{Epidemic processes in complex networks}.
\newblock \emph{\bibinfo{journal}{Rev. Mod. Phys.}}
  \textbf{\bibinfo{volume}{87}}, \bibinfo{pages}{925--979}
  (\bibinfo{year}{2015}).

\bibitem{Sattentau:2008}
\bibinfo{author}{Sattentau, Q.}
\newblock \bibinfo{title}{Avoiding the void: cell-to-cell spread of human
  viruses}.
\newblock \emph{\bibinfo{journal}{Nat. Rev. Microbiol.}}
  \textbf{\bibinfo{volume}{6}}, \bibinfo{pages}{815--826}
  (\bibinfo{year}{2008}).
  
\bibitem{DumonteilETAL:2013}
\bibinfo{author}{Dumonteil, E.}, \bibinfo{author}{Majumdar, S.~N.},
  \bibinfo{author}{Rosso, A.} \& \bibinfo{author}{Zoia, A.}
\newblock \bibinfo{title}{Spatial extent of an outbreak in animal epidemics}.
\newblock \emph{\bibinfo{journal}{Proc. Natl. Acad. Sci. USA}}
  \textbf{\bibinfo{volume}{110}}, \bibinfo{pages}{4239--4244}
  (\bibinfo{year}{2013}).

\bibitem{NekovarPruessner:2016}
\bibinfo{author}{Nekovar, S.} \& \bibinfo{author}{Pruessner, G.}
\newblock \bibinfo{title}{A field-theoretic approach to the wiener sausage}.
\newblock \emph{\bibinfo{journal}{J. Stat. Phys.}}
  \textbf{\bibinfo{volume}{163}}, \bibinfo{pages}{604--641}
  (\bibinfo{year}{2016}).

\bibitem{BerezhkovskiiMakhnovskiiSuris:1989}
\bibinfo{author}{Berezhkovskii, A.~M.}, \bibinfo{author}{Makhnovskii, Y.~A.} \&
  \bibinfo{author}{Suris, R.~A.}
\newblock \bibinfo{title}{Wiener sausage volume moments}.
\newblock \emph{\bibinfo{journal}{J. Stat. Phys.}}
  \textbf{\bibinfo{volume}{57}}, \bibinfo{pages}{333--346}
  (\bibinfo{year}{1989}).

\bibitem{Harris:1963}
\bibinfo{author}{Harris, T.~E.}
\newblock \emph{\bibinfo{title}{The Theory of Branching Processes}}
  (\bibinfo{publisher}{Springer-Verlag}, \bibinfo{address}{Berlin, Germany},
  \bibinfo{year}{1963}).

\bibitem{RamolaMajumdarSchehr:2015}
\bibinfo{author}{Ramola, K.}, \bibinfo{author}{Majumdar, S.~N.} \&
  \bibinfo{author}{Schehr, G.}
\newblock \bibinfo{title}{Spatial extent of branching brownian motion}.
\newblock \emph{\bibinfo{journal}{Phys. Rev. E}} \textbf{\bibinfo{volume}{91}},
  \bibinfo{pages}{042131} (\bibinfo{year}{2015}).

\bibitem{sawyer1979}
\bibinfo{author}{Sawyer, S.} \& \bibinfo{author}{Fleischman, J.}
\newblock \bibinfo{title}{Maximum geographic range of a mutant allele
  considered as a subtype of a brownian branching random field}.
\newblock \emph{\bibinfo{journal}{Proc. Natl. Acad. Sci. USA}}
  \textbf{\bibinfo{volume}{76}}, \bibinfo{pages}{872--875}
  (\bibinfo{year}{1979}).

\bibitem{LeGallLin:2015}
\bibinfo{author}{Le~Gall, J.-F.} \& \bibinfo{author}{Lin, S.}
\newblock \bibinfo{title}{The range of tree-indexed random walk in low
  dimensions}.
\newblock \emph{\bibinfo{journal}{Ann. Probab.}} \textbf{\bibinfo{volume}{43}},
  \bibinfo{pages}{2701--2728} (\bibinfo{year}{2015}).

\bibitem{LeGallLin:2016}
\bibinfo{author}{Le~Gall, J.-F.} \& \bibinfo{author}{Lin, S.}
\newblock \bibinfo{title}{The range of tree-indexed random walk}.
\newblock \emph{\bibinfo{journal}{J. Inst. Math. Jussieu}}
  \textbf{\bibinfo{volume}{15}}, \bibinfo{pages}{271--317}
  (\bibinfo{year}{2016}).


\bibitem{GrimmGrimm:1857}
\bibinfo{author}{Grimm, J.} \& \bibinfo{author}{Grimm, W.}
\newblock \bibinfo{title}{H\"ansel und Grethel}.
\newblock In \emph{\bibinfo{booktitle}{Kinder- und Hausm\"archen}},
  vol.~\bibinfo{volume}{1}, chap.~\bibinfo{chapter}{15},
  \bibinfo{pages}{79--87} (\bibinfo{publisher}{Verlag der Dieterichschen
  Buchhandlung}, \bibinfo{address}{G\"ottingen}, \bibinfo{year}{1857}),
  \bibinfo{edition}{7} edn.

\bibitem{Doi:1976}
\bibinfo{author}{Doi, M.}
\newblock \bibinfo{title}{Second quantization representation for classical
  many-particle system}.
\newblock \emph{\bibinfo{journal}{J. Phys. A: Math. Gen.}}
  \textbf{\bibinfo{volume}{9}}, \bibinfo{pages}{1465--1477}
  (\bibinfo{year}{1976}).

\bibitem{Peliti:1985}
\bibinfo{author}{Peliti, L.}
\newblock \bibinfo{title}{Path integral approach to birth-death processes on a
  lattice}.
\newblock \emph{\bibinfo{journal}{J. Phys. (Paris)}}
  \textbf{\bibinfo{volume}{46}}, \bibinfo{pages}{1469--1483}
  (\bibinfo{year}{1985}).

\bibitem{PfeutyToulouse:1977}
\bibinfo{author}{Pfeuty, P.} \& \bibinfo{author}{Toulouse, G.}
\newblock \emph{\bibinfo{title}{Introduction to the Renormalization Group and
  to Critical Phenomena}} (\bibinfo{publisher}{John Wiley \& Sons},
  \bibinfo{address}{Chichester, West Sussex, UK}, \bibinfo{year}{1977}).

\bibitem{burioni2005}
\bibinfo{author}{Burioni, R.} \& \bibinfo{author}{Cassi, D.}
\newblock \bibinfo{title}{Random walks on graphs: ideas, techniques and
  results}.
\newblock \emph{\bibinfo{journal}{J. Phys. A: Math. Gen.}}
  \textbf{\bibinfo{volume}{38}}, \bibinfo{pages}{R45--R78}
  (\bibinfo{year}{2005}).

\bibitem{barabasi2000}
\bibinfo{author}{Barab{\'a}si, A.-L.}, \bibinfo{author}{Albert, R.} \&
  \bibinfo{author}{Jeong, H.}
\newblock \bibinfo{title}{Scale-free characteristics of random networks: the
  topology of the world-wide web}.
\newblock \emph{\bibinfo{journal}{Physica A}} \textbf{\bibinfo{volume}{281}},
  \bibinfo{pages}{69--77} (\bibinfo{year}{2000}).

\bibitem{albert2005}
\bibinfo{author}{Albert, R.}
\newblock \bibinfo{title}{Scale-free networks in cell biology}.
\newblock \emph{\bibinfo{journal}{J. Cell Sci.}}
  \textbf{\bibinfo{volume}{118}}, \bibinfo{pages}{4947--4957}
  (\bibinfo{year}{2005}).

\bibitem{wang2017}
\bibinfo{author}{Wang, W.}, \bibinfo{author}{Tang, M.},
  \bibinfo{author}{Stanley, H.~E.} \& \bibinfo{author}{Braunstein, L.~A.}
\newblock \bibinfo{title}{Unification of theoretical approaches for epidemic
  spreading on complex networks}.
\newblock \emph{\bibinfo{journal}{Rep. Prog. Phys.}}
  \textbf{\bibinfo{volume}{80}}, \bibinfo{pages}{036603}
  (\bibinfo{year}{2017}).

\bibitem{yu2001some}
\bibinfo{author}{Yu, B.} \& \bibinfo{author}{Li, J.}
\newblock \bibinfo{title}{Some fractal characters of porous media}.
\newblock \emph{\bibinfo{journal}{Fractals}} \textbf{\bibinfo{volume}{9}},
  \bibinfo{pages}{365--372} (\bibinfo{year}{2001}).

\bibitem{lyons1990random}
\bibinfo{author}{Lyons, R.}
\newblock \bibinfo{title}{Random walks and percolation on trees}.
\newblock \emph{\bibinfo{journal}{Ann. Probab.}} \bibinfo{pages}{931--958}
  (\bibinfo{year}{1990}).

\bibitem{watanabe1985spectral}
\bibinfo{author}{Watanabe, H.}
\newblock \bibinfo{title}{Spectral dimension of a wire network}.
\newblock \emph{\bibinfo{journal}{J. Phys. A: Math. Gen.}}
  \textbf{\bibinfo{volume}{18}}, \bibinfo{pages}{2807--2823}
  (\bibinfo{year}{1985}).

\bibitem{destri2002}
\bibinfo{author}{Destri, C.} \& \bibinfo{author}{Donetti, L.}
\newblock \bibinfo{title}{The spectral dimension of random trees}.
\newblock \emph{\bibinfo{journal}{J. Phys. A: Math. Gen.}}
  \textbf{\bibinfo{volume}{35}}, \bibinfo{pages}{9499–-9515}
  (\bibinfo{year}{2002}).

\bibitem{barabasi1999emergence}
\bibinfo{author}{Barab{\'a}si, A.-L.} \& \bibinfo{author}{Albert, R.}
\newblock \bibinfo{title}{Emergence of scaling in random networks}.
\newblock \emph{\bibinfo{journal}{Science}} \textbf{\bibinfo{volume}{286}},
  \bibinfo{pages}{509--512} (\bibinfo{year}{1999}).

\bibitem{barabasi2000scale}
\bibinfo{author}{Barab{\'a}si, A.-L.}, \bibinfo{author}{Albert, R.} \&
  \bibinfo{author}{Jeong, H.}
\newblock \bibinfo{title}{Scale-free characteristics of random networks: the
  topology of the world-wide web}.
\newblock \emph{\bibinfo{journal}{Physica A}} \textbf{\bibinfo{volume}{281}},
  \bibinfo{pages}{69--77} (\bibinfo{year}{2000}).

\bibitem{guimera2005worldwide}
\bibinfo{author}{Guimera, R.}, \bibinfo{author}{Mossa, S.},
  \bibinfo{author}{Turtschi, A.} \& \bibinfo{author}{Amaral, L.~N.}
\newblock \bibinfo{title}{The worldwide air transportation network: Anomalous
  centrality, community structure, and cities' global roles}.
\newblock \emph{\bibinfo{journal}{Proc. Natl. Acad. Sci. USA}}
  \textbf{\bibinfo{volume}{102}}, \bibinfo{pages}{7794--7799}
  (\bibinfo{year}{2005}).

\bibitem{jeong2000large}
\bibinfo{author}{Jeong, H.}, \bibinfo{author}{Tombor, B.},
  \bibinfo{author}{Albert, R.}, \bibinfo{author}{Oltvai, Z.~N.} \&
  \bibinfo{author}{Barab{\'a}si, A.-L.}
\newblock \bibinfo{title}{The large-scale organization of metabolic networks}.
\newblock \emph{\bibinfo{journal}{Nature}} \textbf{\bibinfo{volume}{407}},
  \bibinfo{pages}{651} (\bibinfo{year}{2000}).

\bibitem{samukhin2008laplacian}
\bibinfo{author}{Samukhin, A.~N.}, \bibinfo{author}{Dorogovtsev, S.~N.} \&
  \bibinfo{author}{Mendes, J. F.~F.}
\newblock \bibinfo{title}{Laplacian spectra of, and random walks on, complex
  networks: Are scale-free architectures really important?}
\newblock \emph{\bibinfo{journal}{Phys. Rev. E}} \textbf{\bibinfo{volume}{77}},
  \bibinfo{pages}{036115} (\bibinfo{year}{2008}).

\bibitem{masuda2017}
\bibinfo{author}{Masuda, N.}, \bibinfo{author}{Porter, M.~A.} \&
  \bibinfo{author}{Lambiotte, R.}
\newblock \bibinfo{title}{Random walks and diffusion on networks}.
\newblock \emph{\bibinfo{journal}{Phys. Rep.}} \textbf{\bibinfo{volume}{716}},
  \bibinfo{pages}{1--58} (\bibinfo{year}{2017}).

\bibitem{viswanath-2009-activity}
\bibinfo{author}{Viswanath, B.}, \bibinfo{author}{Mislove, A.},
  \bibinfo{author}{Cha, M.} \& \bibinfo{author}{Gummadi, K.~P.}
\newblock \bibinfo{title}{On the evolution of user interaction in facebook}.
\newblock In \emph{\bibinfo{booktitle}{Proceedings of the 2nd ACM workshop on
  Online social networks}}, \bibinfo{pages}{37--42}
  (\bibinfo{organization}{ACM}, \bibinfo{year}{2009}).

\bibitem{jeong2001lethality}
\bibinfo{author}{Jeong, H.}, \bibinfo{author}{Mason, S.~P.},
  \bibinfo{author}{Barab{\'a}si, A.-L.} \& \bibinfo{author}{Oltvai, Z.~N.}
\newblock \bibinfo{title}{Lethality and centrality in protein networks}.
\newblock \emph{\bibinfo{journal}{Nature}} \textbf{\bibinfo{volume}{411}},
  \bibinfo{pages}{41--42} (\bibinfo{year}{2001}).

\bibitem{gallos2007scaling}
\bibinfo{author}{Gallos, L.~K.}, \bibinfo{author}{Song, C.},
  \bibinfo{author}{Havlin, S.} \& \bibinfo{author}{Makse, H.~A.}
\newblock \bibinfo{title}{Scaling theory of transport in complex biological
  networks}.
\newblock \emph{\bibinfo{journal}{Proc. Natl. Acad. Sci. USA}}
  \textbf{\bibinfo{volume}{104}}, \bibinfo{pages}{7746--7751}
  (\bibinfo{year}{2007}).

\bibitem{han2005effect}
\bibinfo{author}{Han, J.-D.~J.}, \bibinfo{author}{Dupuy, D.},
  \bibinfo{author}{Bertin, N.}, \bibinfo{author}{Cusick, M.~E.} \&
  \bibinfo{author}{Vidal, M.}
\newblock \bibinfo{title}{Effect of sampling on topology predictions of
  protein-protein interaction networks}.
\newblock \emph{\bibinfo{journal}{Nat. Biotechnol.}}
  \textbf{\bibinfo{volume}{23}}, \bibinfo{pages}{839–-844}
  (\bibinfo{year}{2005}).

\bibitem{stumpf2005subnets}
\bibinfo{author}{Stumpf, M.~P.}, \bibinfo{author}{Wiuf, C.} \&
  \bibinfo{author}{May, R.~M.}
\newblock \bibinfo{title}{Subnets of scale-free networks are not scale-free:
  sampling properties of networks}.
\newblock \emph{\bibinfo{journal}{Proc. Natl. Acad. Sci. USA}}
  \textbf{\bibinfo{volume}{102}}, \bibinfo{pages}{4221--4224}
  (\bibinfo{year}{2005}).

\bibitem{cardy2008non}
\bibinfo{author}{Cardy, J.}, \bibinfo{author}{Falkovich, G.} \&
  \bibinfo{author}{Gaw{\k{e}}dzki, K.}
\newblock \emph{\bibinfo{title}{Non-equilibrium statistical mechanics and
  turbulence}}, vol. \bibinfo{volume}{355} (\bibinfo{publisher}{Cambridge
  University Press}, \bibinfo{year}{2008}).

\bibitem{simonsen2004}
\bibinfo{author}{Simonsen, I.}, \bibinfo{author}{Eriksen, K.~A.},
  \bibinfo{author}{Maslov, S.} \& \bibinfo{author}{Sneppen, K.}
\newblock \bibinfo{title}{Diffusion on complex networks: a way to probe their
  large-scale topological structures}.
\newblock \emph{\bibinfo{journal}{Physica A}} \textbf{\bibinfo{volume}{336}},
  \bibinfo{pages}{163--173} (\bibinfo{year}{2004}).

\bibitem{TaeuberHowardVollmayr-Lee:2005}
\bibinfo{author}{T{\"a}uber, U.~C.}, \bibinfo{author}{Howard, M.} \&
  \bibinfo{author}{Vollmayr-Lee, B.~P.}
\newblock \bibinfo{title}{Applications of field-theoretic renormalization group
  methods to reaction-diffusion problems}.
\newblock \emph{\bibinfo{journal}{J. Phys. A: Math. Gen.}}
  \textbf{\bibinfo{volume}{38}}, \bibinfo{pages}{R79--R131}
  (\bibinfo{year}{2005}).

\bibitem{Taeuber:2014}
\bibinfo{author}{T{\"a}uber, U.~C.}
\newblock \emph{\bibinfo{title}{Critical Dynamics A Field Theory Approach to
  Equilibrium and Non-Equilibrium Scaling Behavior}}
  (\bibinfo{publisher}{Cambridge University Press},
  \bibinfo{address}{Cambridge, England}, \bibinfo{year}{2014}).

\bibitem{dasgupta1999scaling}
\bibinfo{author}{Dasgupta, R.}, \bibinfo{author}{Ballabh, T.~K.} \&
  \bibinfo{author}{Tarafdar, S.}
\newblock \bibinfo{title}{Scaling exponents for random walks on sierpinski
  carpets and number of distinct sites visited: a new algorithm for infinite
  fractal lattices}.
\newblock \emph{\bibinfo{journal}{J. Phys. A: Math. Gen.}}
  \textbf{\bibinfo{volume}{32}}, \bibinfo{pages}{6503--6516}
  (\bibinfo{year}{1999}).

\bibitem{destri2002growth}
\bibinfo{author}{Destri, C.} \& \bibinfo{author}{Donetti, L.}
\newblock \bibinfo{title}{On the growth of bounded trees}.
\newblock \emph{\bibinfo{journal}{J. Phys. A: Math. Gen.}}
  \textbf{\bibinfo{volume}{35}}, \bibinfo{pages}{5147} (\bibinfo{year}{2002}).

\end{thebibliography}

\begin{thebibliography}{10}
\expandafter\ifx\csname url\endcsname\relax
  \def\url#1{\texttt{#1}}\fi
\expandafter\ifx\csname urlprefix\endcsname\relax\def\urlprefix{URL }\fi
\providecommand{\bibinfo}[2]{#2}
\providecommand{\eprint}[2][]{\url{#2}}

\bibitem{SNekovarPruessner:2016}
\bibinfo{author}{Nekovar, S.} \& \bibinfo{author}{Pruessner, G.}
\newblock \bibinfo{title}{A field-theoretic approach to the wiener sausage}.
\newblock \emph{\bibinfo{journal}{J. Stat. Phys.}}
  \textbf{\bibinfo{volume}{163}}, \bibinfo{pages}{604--641}
  (\bibinfo{year}{2016}).

\bibitem{STaeuber:2014}
\bibinfo{author}{T{\"a}uber, U.~C.}
\newblock \emph{\bibinfo{title}{Critical Dynamics A Field Theory Approach to
  Equilibrium and Non-Equilibrium Scaling Behavior}}
  (\bibinfo{publisher}{Cambridge University Press},
  \bibinfo{address}{Cambridge, England}, \bibinfo{year}{2014}).

\bibitem{SLeBellac:1991}
\bibinfo{author}{{Le Bellac}, M.}
\newblock \emph{\bibinfo{title}{Quantum and Statistical Field Theory
  [Phenomenes critiques aux champs de jauge, English]}}
  (\bibinfo{publisher}{Oxford University Press}, \bibinfo{address}{New York,
  NY, USA}, \bibinfo{year}{1991}).
\newblock \bibinfo{note}{Translated by G. Barton}.

\bibitem{Saldous1993continuum}
\bibinfo{author}{Aldous, D.}
\newblock \bibinfo{title}{The continuum random tree iii}.
\newblock \emph{\bibinfo{journal}{Ann. Probab.}} \textbf{\bibinfo{volume}{21}},
  \bibinfo{pages}{248--289} (\bibinfo{year}{1993}).

\bibitem{Sle2005random}
\bibinfo{author}{Le~Gall, J.-F.}
\newblock \bibinfo{title}{Random trees and applications}.
\newblock \emph{\bibinfo{journal}{Probab. Surv.}} \textbf{\bibinfo{volume}{2}},
  \bibinfo{pages}{245--311} (\bibinfo{year}{2005}).

\bibitem{SGarcia-MillanETAL:2018}
\bibinfo{author}{Garcia-Millan, R.}, \bibinfo{author}{Pausch, J.},
  \bibinfo{author}{Walter, B.} \& \bibinfo{author}{Pruessner, G.}
\newblock \bibinfo{title}{Field-theoretic approach to the universality of
  branching processes}.
\newblock \emph{\bibinfo{journal}{Phys. Rev. E}} \textbf{\bibinfo{volume}{98}},
  \bibinfo{pages}{062107} (\bibinfo{year}{2018}).

\bibitem{Sburioni2005}
\bibinfo{author}{Burioni, R.} \& \bibinfo{author}{Cassi, D.}
\newblock \bibinfo{title}{Random walks on graphs: ideas, techniques and
  results}.
\newblock \emph{\bibinfo{journal}{J. Phys. A: Math. Gen.}}
  \textbf{\bibinfo{volume}{38}}, \bibinfo{pages}{R45--R78}
  (\bibinfo{year}{2005}).

\bibitem{SDiehlSchmidt:2011}
\bibinfo{author}{Diehl, H.~W.} \& \bibinfo{author}{Schmidt, F.~M.}
\newblock \bibinfo{title}{The critical casimir effect in films for generic
  non-symmetry-breaking boundary conditions}.
\newblock \emph{\bibinfo{journal}{New J. Phys.}} \textbf{\bibinfo{volume}{13}},
  \bibinfo{pages}{123025} (\bibinfo{year}{2011}).

\bibitem{SPressETAL:1992}
\bibinfo{author}{Press, W.~H.}, \bibinfo{author}{Teukolsky, S.~A.},
  \bibinfo{author}{Vetterling, W.~T.} \& \bibinfo{author}{Flannery, B.~P.}
\newblock \emph{\bibinfo{title}{Numerical Recipes in C}}
  (\bibinfo{publisher}{Cambridge University Press}, \bibinfo{address}{New York,
  NY, USA}, \bibinfo{year}{1992}), \bibinfo{edition}{2nd} edn. 
\end{thebibliography}
\end{document}